\documentclass{article}
\usepackage[a4paper, total={6in, 8in}]{geometry}
\usepackage[T1]{fontenc}
\usepackage[utf8]{inputenc}
\usepackage{authblk}

\usepackage{mathtools} 
\usepackage{amsmath}
\usepackage{braket}
\usepackage{graphicx}
\usepackage{caption}
\usepackage{subcaption}
\usepackage[english]{babel}
\usepackage{amssymb}
\usepackage{xcolor}
\usepackage{amsfonts}

\usepackage{cite}

\usepackage{indentfirst}

\title{Entanglement swapping and swapped entanglement}
\def\correspondingauthor{\footnote{Corresponding author}}
\author[1]{S. M. Zangi \correspondingauthor{}}
\author[2]{Chitra Shukla}
\author[3]{Atta Ur Rahman}
\author[4]{Bo Zheng \correspondingauthor{}}
\affil[1,4]{School of Physics and Astronomy and Yunnan Key Laboratory for Quantum Information, Yunnan University, Kunming 650500, China}
\affil[2]{Shenzhen Institute for Quantum Science and Engineering and               Department of Physics, Southern University of Science and Technology, Shenzhen 518055, China}
\affil[3]{School of Physics, University of Chinese Academy of Science,
          Yuquan Road 19A, Beijing 100049, China}
\date{}

\begin{document}

    \maketitle
    
    \begin{abstract}
        Entanglement swapping is gaining widespread attention due to its application in entanglement distribution among different parts of quantum appliances. We investigate the entanglement swapping for pure and noisy systems, and argue different  entanglement quantifiers for quantum states. We explore the relationship between the entanglement of initial states  and the average entanglement of final states in terms of concurrence and negativity. We find that if initial quantum states are maximally entangled and we make measurements in the Bell basis, then average concurrence and average negativity of final states give similar results. In this case, we simply obtain the average concurrence (average negativity) of the final states by taking the product of concurrences (negativities) of the initial states. However, the measurement in non-maximally entangled basis during entanglement swapping degrades the average swapped entanglement. Further, the product of the entanglement of the initial mixed states provides an upper bound to the average swapped entanglement of final states obtained after entanglement swapping. The average concurrence of finally obtained states provides an upper bound to the average negativity of these states. We also discuss how successfully the output state can be used as a channel for the teleportation of an unknown qubit.
    \end{abstract}
    
    \section{Introduction}
        Quantum networks play a very important role in quantum information science and have applications in quantum
        communication \cite{qiu2014quantum}, computation \cite{PhysRevA.98.030302}, metrology \cite{chiribella2012optimal} and fundamental tests \cite{mccutcheon2016experimental}. Quantum networks are comprised of a large number of nodes interconnected by quantum channels. The stationary qubits at the separated nodes develop, store, and manipulate quantum states while the flying qubits constitute quantum channels and can be realized by photons. These channels teleport quantum states between the nodes with high fidelity, allowing the distribution of quantum entanglement across the whole network. Thus, the task of building quantum networks requires the ability to establish quantum entanglement between distant quantum nodes. H. J. Briegel et al. proposed a quantum repeater protocol that enables the transmission of entanglement over long distances \cite{briegel1998quantum}. The ability to distribute and manipulate entanglement between distant parties serves as the basis for quantum applications. Entanglement swapping is one such tool that helps us to connect many separable nodes for long-distance communication in a quantum network. Specifically, entanglement swapping is a protocol by which quantum systems that have never interacted in the past can become entangled \cite{PhysRevLett.71.4287,PhysRevA.72.042310}. The nomenclature “entanglement swapping” describes the transfer of entanglement from a priori entangled systems to a priori separable systems \cite{vedral2006introduction}.
        It is a very useful tool for entanglement purification \cite{ji2022entanglement}, teleportation \cite{PhysRevLett.70.1895}, and plays an important role in quantum computing and quantum cryptography \cite{galindo2002information,ji2019quantum}. We can also use entanglement swapping for the creation of multipartite entangled states from bipartite entanglement \cite{PhysRevA.57.822}. 
        
\begin{figure}
            \centering
            \begin{subfigure}[t]{0.49\textwidth}
            \centering
            \includegraphics[width=\textwidth]{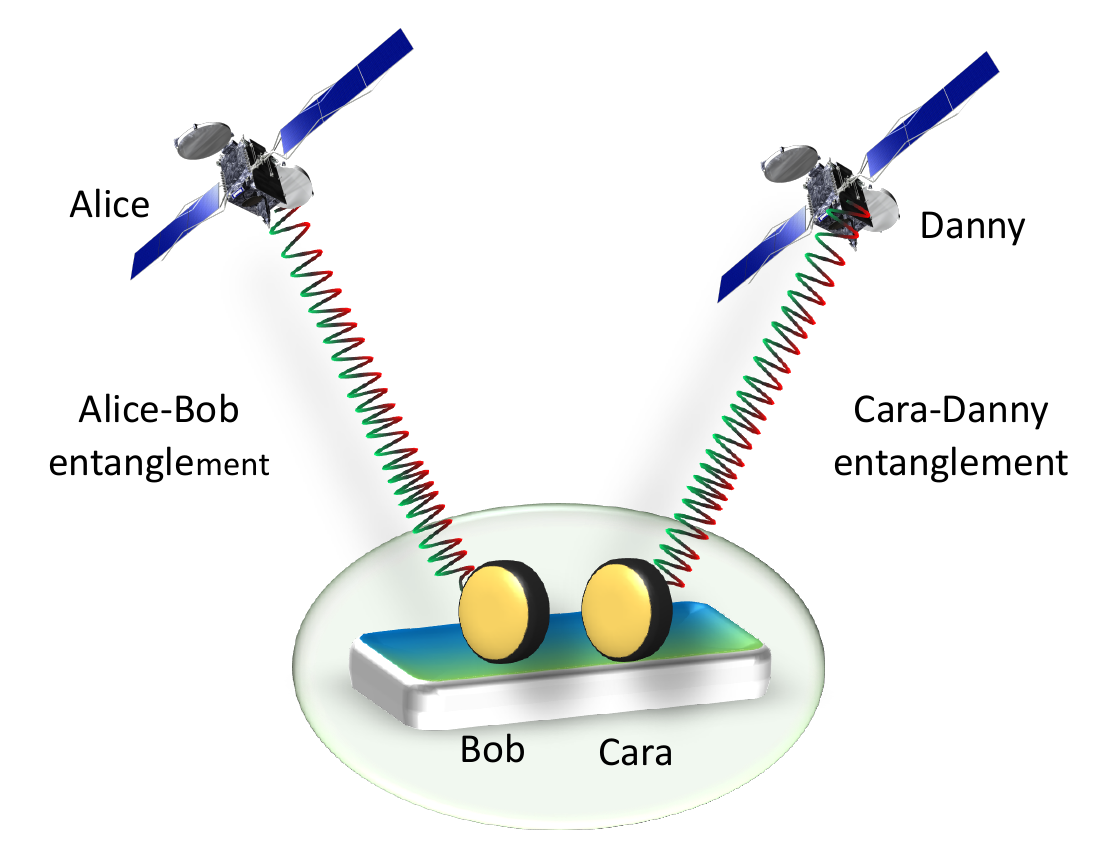}
            \caption{}
            \label{fig:b-measurement}
            \end{subfigure}
            \hfill
            \begin{subfigure}[t]{0.49\textwidth}
            \centering
            \includegraphics[width=\textwidth]{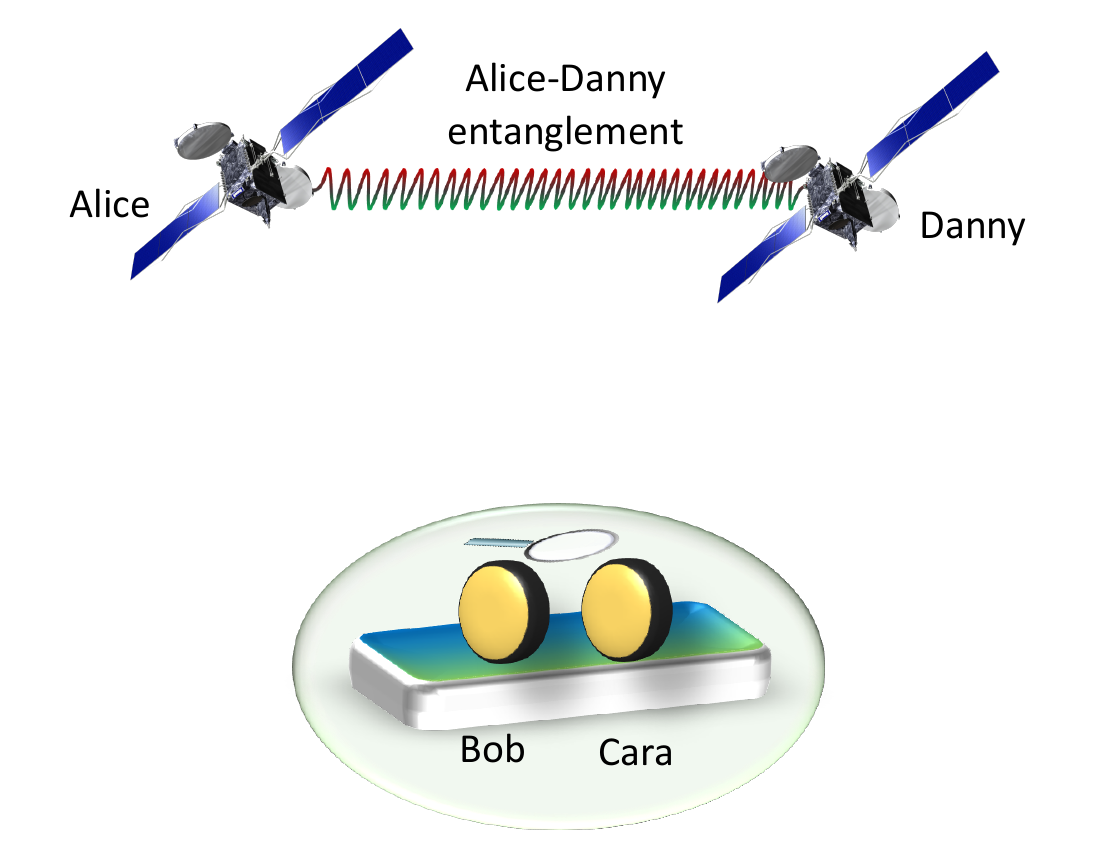}
            \caption{}
            \label{fig:a-measurement}
            \end{subfigure}
            \hfill
     
            \caption{Entanglement swapping. Initially in Fig. (a)  entangled pairs are shared between Alice and Bob, and between Cara and Danny. There is no entanglement between Alice and Danny. However, in Fig (b), the measurement on Bob and Cara's qubits project the entanglement between Alice and Danny.}
            \label{fig:e-swapping}
        \end{figure}

        Let us describe the phenomenon of entanglement swapping. Suppose two entangled particles $(A, B)$ are shared between Alice and Bob. Similarly Cara and Danny also share another entangled pair of particles $(C, D)$. Initially there is no entanglement between Alice’s and Danny’s particles $(A, D)$, shown in Fig. (\ref{fig:b-measurement}). If Bob and Cara who are situated in the same laboratory make measurement in a suitable basis on the pair $(B, C)$ and classically communicate the outcome with distant partners then Alice’s and Danny’s particles who are at very large distance become entangled as shone in Fig. (\ref{fig:a-measurement}). This entanglement swapping protocol can be generalized in different ways: by modifying the initial states, or by modifying the measurement performed by Bob and Cara, or by extending the number of parties \cite{PhysRevA.57.822,PhysRevLett.93.260501}.

        An ever-increasing body of literature shows that the entanglement swapping and purification of quantum systems need specific protocols. Smaller changes may bring huge changes to the output state because of the relative sensitivity of the operation and quantum systems. For this reason, several previous studies suggest entanglement swapping of initial states into maximally entangled states. For example, in Ref. \cite{song2014purifying}, the authors provided a scheme of entanglement swapping of initial states into biqubit maximally entangled states when influenced by an amplitude damping channel. The concurrence of the measuring basis for entanglement swapping caused a two-fold entanglement matching effect has been witnessed in Ref. \cite{roa2021matching}. Recently, Ref. \cite{huang2022entanglement} investigated that hyper-entangled states produce deterministic entanglement swapping while considering projections of biqubit systems on symmetric, and iso-entangled states. It is found that biqubit entanglement generated through entanglement swapping, will depart from a Bell-type inequality even for visibilities smaller than 50\% \cite{branciard2010characterizing}.  From the above literature, we found that entanglement swapping has been previously considered using different procedures, and various important results have been achieved. This research work constitutes a relatively more generalized entanglement swapping protocol that covers the maximum possible cases of swapping. We consider pure, mixed as well as noisy systems for entanglement swapping. Moreover, we explore the application of the final state as a channel for the teleportation of an unknown qubit. We also investigate the fidelity of teleported qubit with the initial unknown qubit.
        
        The article is organized as follows. In Section \ref{Preliminaries}, we present the mathematical tools that are useful for the present research. The section \ref{esaqs} describes the details of the entanglement swapping scenario for non-noisy systems and the application of swapped entanglement. Next, in Section \ref{nqaes}, we demonstrate the entanglement swapping among noisy qubits and teleportation using a noisy quantum channel. Finally, we conclude with a short discussion in Section \ref{cd}.

    \section{Preliminaries}\label{Preliminaries}
        In this section, we explain in brief the Schmidt decomposition, entangled quantum states and famous entanglement quantifiers. We shall quantify the swapped entanglement with these quantifiers and compare them in next sections. 
        
        \subsection{Schmidt  decomposition}
             Suppose $\ket{\psi}_{XY}$ is a pure state of a composite system $XY$. According to Schmidt decomposition \cite{nielsen2010quantum} there exist orthonormal basis $\ket{i}_{X}$ for system X, and orthonormal basis $\ket{i}_{Y}$ of system Y such that 
        \begin{equation}\label{eq1}
             \ket{\psi}=\Sigma_{i}\sqrt{p_{i}}\ket{i}_{X}\ket{i}_{Y}\;,  
        \end{equation}
        where $\sqrt{p_{i}}$ are non-negative real numbers satisfying $\Sigma_{i}p_{i}=1$ and known as Schmidt coefficients.
        \subsection{Entangled states}
            A quantum state of a compound system is entangled if it cannot be written as a convex combination of product
            states. A bipartite quantum state represented by product states $\rho_i^X\otimes \rho_i^Y$ with convex weights $p_i$ as 
            \begin{equation}\label{eq2}
                \rho_{XY}=\sum_{i}^{L}p_i\rho_i^X\otimes \rho_i^Y\;,
            \end{equation}
            is called a separable state \cite{zangi2021combo}.
        \subsection{Entanglement quantifiers}
            Many entanglement quantifiers have been proposed \cite{bennett1996mixed,PhysRevA.57.1619,PhysRevA.75.052320,lee2003convex}, however concurrence and negativity are among the well-defined measures of entanglement \cite{karimi2019measurability,PhysRevA.65.032314,PhysRevA.91.032327,miranowicz2004comparative}.
            
           \subsubsection{Concurrence}
            For a finite-dimensional bipartite state $\ket{\psi}\in H_{X}\otimes H_{Y}$, where $H_{X}(H_{Y})$ denotes the $M(N)$-dimensional vector space associated with the subsystem $X(Y)$ such that $M \leq N$, the concurrence is
               \begin{equation}\label{eq3}
                    \textbf{C}(\psi) = \sqrt{2\left(1-\operatorname{Tr}\left(\rho_{X}^{2}\right)\right)}\;,
                \end{equation}
            where the reduced density operator $\rho_{X}=Tr_{Y}(\ket{\psi}\bra{\psi})$. Sometimes for $n-$dimensional system, the factor 2 is replaced by $\frac{n}{n-1}$, nonetheless this difference in normalization is not essential. In order to compute the concurrence from the Schmidt decomposition of biqubit state $\ket{\psi}$, we can use the following simple relation:
                \begin{equation}\label{eq4}
                  \textbf{C}(\psi) =2\sqrt{\sum_{j< k}p_{j} p_{k}} \;.  
                \end{equation}
            The above equation implies that concurrence is the root mean square of pair-wise products of the Schmidt coefficients.
                
           \subsubsection{Negativity}
                Entanglement of a bipartite state $\rho_{XY}$ can also be quantified by using the concept of negativity defined as
                \begin{equation} \label{neg}
                    \mathcal{N}(\rho)=\frac{\left\|\rho^{T_{Y}}\right\|_{1}-1}{M-1}
                \end{equation}
    
                where $\rho^{T_Y}$ is partial transpose  of the total state $\rho_{XY}$ with respect to the subsystem $Y$, defined as $\left\langle i_{X}, j_{Y}\left|\rho^{T_B}\right| k_{X}, l_{Y}\right\rangle=\left\langle k_{X}, j_{Y}\left|\rho\right| i_{X}, l_{Y}\right\rangle$. The huge advantage of the negativity is that it can easily be computed even for a mixed state.
                
                If a biqubit state $\ket{\psi}$ given in the Schmidt decomposition form then  the negativity can be computed by taking the sum of pair-wise products of the Schmidt coefficients as
                
                \begin{equation}
                    \mathcal{N}(\psi)=2 \sum_{j< k}\sqrt{p_{j}}\sqrt{p_{k}} \;. 
                \end{equation}
                Here, we use the convention given in Ref. \cite{sabin2008classification} for the measurement of negativity, which is twice the value of the original definition. It is important to note that $C\geq \mathcal{N}$ \cite{rai2005negativity}. The maximum value of $C$ and $\mathcal{N}$ is one for maximally entangled states.
                
    \section{Entanglement swapping among qubit systems}\label{esaqs}
        We consider two pairs of qubits for entanglement swapping. The entangled qubits $A$ and $B$ make the first pair and the second pair consist of entangled qubits $C$ and $D$ (the qubits' names are chosen such that they make the initial of Alice, Bob, Cara and Danny, shown in Fig.(\ref{fig:e-swapping})). In terms of Schmidt decomposition, these subsystems can be written as
    
        \begin{equation}\label{eq5}
        \begin{split}
            \ket{\phi}_{AB} &=\sqrt{p_{0}}\ket{00}_{AB}+\sqrt{p_{1}}\ket{11}_{AB}, \\
            \ket{\phi}_{CD} &=\sqrt{p_{0}^\prime}\ket{00}_{CD}+\sqrt{p_{1}^\prime}\ket{11}_{CD}\;.
        \end{split}
        \end{equation}
        The concurrence and negativity of these systems are $\textbf{C}(\ket{\phi}_{AB})=2\sqrt{p_{0}p_{1}}$, $\textbf{C}(\ket{\phi}_{CD})=2\sqrt{p_{0}^\prime p_{1}^\prime}$ and  $\mathcal{N}(\ket{\phi}_{AB})=2\sqrt{p_{0}}\sqrt{p_{1}}$, $\mathcal{N}(\ket{\phi}_{CD})=2\sqrt{p_{0}^\prime}\sqrt{ p_{1}^\prime}$. It means for these bi-dimensional systems concurrence and negativity produce similar results. The initial state of our four-qubit system can be written as  
        \begin{equation}\label{eq6}
        \begin{split}
             \ket{\Phi}&= \ket{\phi}_{AB}\ket{\phi}_{CD}\\
            &=\sqrt{p_{0}p_{0}^\prime}\ket{0000}_{ABCD}+
            \sqrt{p_{0}p_{1}^\prime}\ket{0011}_{ABCD}+
            \sqrt{p_{1}p_{0}^\prime}\ket{1100}_{ABCD}+
            \sqrt{p_{1}p_{1}^\prime}\ket{1111}_{ABCD}\;.
        \end{split}
        \end{equation}
        Let us do some simple algebra. We rearrange the qubits in each term so that we can write qubits $A$ and $D$ together and qubits $B$ and $C$ together: 
        \begin{equation}\label{eq7}
             \ket{\Phi}=
            \sqrt{p_{0}p_{0}^\prime}\ket{00}_{AD}\ket{00}_{BC}
            \sqrt{p_{0}p_{1}^\prime}\ket{01}_{AD}\ket{01}_{BC}+
            \sqrt{p_{1}p_{0}^\prime}\ket{10}_{AD}\ket{10}_{BC}+
            \sqrt{p_{1}p_{1}^\prime}\ket{11}_{AD}\ket{11}_{BC}\;.
        \end{equation}
        In order to do measurements over $BC$ qubits, we also need to define a set of four orthonormal basis 
        \cite{roa2021matching}
        \begin{equation}\label{eq10}
        \begin{split}
             \ket{\tilde{\Phi}_{+}}_{BC}&=\alpha_{0}\ket{00}_{BC}+\beta_{0}\ket{11}_{BC},\\
             \ket{\tilde{\Phi}_{-}}_{BC}&=\beta_{0}^{*}\ket{00}_{BC}-\alpha_{0}^{*}\ket{11}_{BC},\\
             \ket{\tilde{\Psi}_{+}}_{BC}&=\alpha_{1}\ket{01}_{BC}+\beta_{1}\ket{10}_{BC},\\
             \ket{\tilde{\Psi}_{-}}_{BC}&=\beta_{1}^{*}\ket{01}_{BC}-\alpha_{1}^{*}\ket{10}_{BC}\;,
        \end{split}
        \end{equation}
        where, without loss of generality, we assume the amplitudes $\alpha_{i}$ and $\beta_{i}$ to be real and non-negative numbers and for normalization $\lvert \alpha_{i}\rvert^2+\lvert\beta_{i}\rvert^2 = 1$ for $i\in\{0,1\}$. Conversely, we have $\ket{00}_{BC} = \alpha_{0}^{*}\ket{\tilde{\Phi}_{+}}_{BC}+ \beta_{0}\ket{\tilde{\Phi}_{-}}_{BC}$ and likewise we can find expressions for $\ket{01}_{BC}, \ket{10}_{BC}, \ket{11}_{BC}$. Now by using these expressions we can write Eq. \eqref{eq7} as
        \begin{multline}\label{eq11}
             \ket{\Phi}=\sqrt{p_{\tilde{\Phi}_{+}}}\ket{\ddot{\Phi}_{+}}_{AD} \ket{\tilde{\Phi}_{+}}_{BC}
             +\sqrt{p_{\tilde{\Phi}_{-}}}\ket{\ddot{\Phi}_{-}}_{AD}\ket{\tilde{\Phi}_{-}}_{BC}\\
             +\sqrt{p_{\tilde{\Psi}_{+}}}\ket{\ddot{\Psi}_{+}}_{AD}\ket{\tilde{\Psi}_{+}}_{BC}
             +\sqrt{p_{\tilde{\Psi}_{-}}}\ket{\ddot{\Psi}_{-}}_{AD}\ket{\tilde{\Psi}_{-}}_{BC}\;,
       \end{multline}
        where possible outcome states of the qubits AD can be defined as,
    \begin{equation}\label{eq:swap2q}
        \begin{split}
             \ket{\ddot{\Phi}_{+}}_{AD}=&\left(\sqrt{p_{0} p_{0}^{\prime}} \alpha_{0}^{*}\ket{0}_{A}\ket{0}_{D}+\sqrt{p_{1} p_{1}^{\prime}} \beta_{0}^{*}\ket{1}_{A}\ket{1}_{D}\right)/\sqrt{p_{\tilde{\Phi}_{+}}}\;,\\
            \ket{\ddot{\Phi}_{-}}_{AD} =&\left(\sqrt{p_{0} p_{0}^{\prime}} \beta_{0}\ket{0}_{A}\ket{0}_{D}-\sqrt{p_{1} p_{1}^{\prime}} \alpha_{0}\ket{1}_{A}\ket{1}_{D}\right)/\sqrt{p_{\tilde{\Phi}_{-}}}\;,\\
            \ket{\ddot{\Psi}_{+}}_{AD} =&\left(\sqrt{p_{0} p_{1}^{\prime}} \alpha_{1}^{*}\ket{0}_{A}\ket{1}_{D}
            +\sqrt{p_{0}^{\prime} p_{1}} \beta_{1}^{*}\ket{1}_{A}\ket{0}_{D}\right)/\sqrt{p_{\tilde{\Psi}_{+}}}\;,\\
        \ket{\ddot{\Psi}_{-}}_{AD}=&\left(\sqrt{p_{0} p_{1}^{\prime}}\beta_{1}\ket{0}_{A}\ket{1}_{D}-\sqrt{p_{0}^{\prime} p_{1}} \alpha_{1}\ket{1}_{A}\ket{0}_{D}\right)/\sqrt{p_{\tilde{\Psi}_{-}}} \;,
    \end{split}
    \end{equation}
    where the associated probabilities are
    \begin{equation}\label{eq13}
        \begin{split}
            p_{\tilde{\Phi}_{+}}&= p_{0} p_{0}^{\prime}\lvert \alpha_{0}\rvert^2 + p_{1} p_{1}^{\prime}\lvert \beta_{0}\rvert^2  \;,\\
            p_{\tilde{\Phi}_{-}}&= p_{0} p_{0}^{\prime}\lvert \beta_{0}\rvert^2 + p_{1} p_{1}^{\prime}\lvert \alpha_{0}\rvert^2 \;,\\
            p_{\tilde{\Psi}_{+}}&= p_{0} p_{1}^{\prime}\lvert \alpha_{1}\rvert^2 + p_{0}^{\prime}p_{1} \lvert \beta_{1}\rvert^2  \;,\\
            p_{\tilde{\Psi}_{-}}&= p_{0} p_{1}^{\prime}\lvert \beta_{1}\rvert^2 + p_{0}^{\prime}p_{1}\lvert \alpha_{1}\rvert^2\;.
    \end{split}
    \end{equation}
    We observe in Eq. (\ref{eq11}) that the state of qubits $A$ and $D$ is similar to the state of the basis $BC$. It is clear from equation (\ref{eq:swap2q}) that after measurements in the basis $BC$, Alice and Danny's qubits $A$, $D$ which are initially separable, become entangled in one of the four possible forms. We can compute the average concurrence for the final state as
    \begin{equation}\label{C-av}
    \begin{split}
    \textbf{C}_{av} &=p_{\tilde{\Phi}_{+}}\textbf{C}_{\ddot{\Phi}_{+}}+p_{\tilde{\Phi}_{-}}\textbf{C}_{\ddot{\Phi}_{-}}+p_{\tilde{\Psi}_{+}}\textbf{C}_{\ddot{\Psi}_{+}}+p_{\tilde{\Psi}_{-}}\textbf{C}_{\ddot{\Psi}_{-}}\;,\\
    &= 4\sqrt{ p_{0} p_{0}^{\prime} p_{1} p_{1}^{\prime}}\left(\lvert\alpha_{0}\beta_{0}\rvert+\lvert\alpha_{1}\beta_{1}\rvert\right)\;.
    \end{split}
    \end{equation}
    
    The Bell states are maximally entangled biqubit states. The states in Eq. (\ref{eq10}) transform into maximally entangled Bell states if we take $\alpha_{i}=\beta_{i}=\frac{1}{\sqrt{2}}$ for $i\in\{0,1\}$. The measurement in maximally entangled Bell basis return maximally entangled $A$, $D$ qubits states with the average concurrence
    \begin{equation}\label{C-mav}
        \begin{split}
            \textbf{C}_{av} &=  4\sqrt{ p_{0} p_{0}^{\prime} p_{1}p_{1}^{\prime}}\;\\
            &= \textbf{C}_{AB}\textbf{C}_{CD}\;.
        \end{split}
    \end{equation}
    Similarly, the average negativity of the qubits $A$ and $D$ states when the measurement is done in Bell basis takes the form
    \begin{equation}\label{N-mav}
        \begin{split}
            \mathcal{N}_{av} &=4\sqrt{ p_{0} p_{0}^{\prime} p_{1}p_{1}^{\prime}}\;\\
            &=\mathcal{N}_{AB}\mathcal{N}_{CD}\;.
        \end{split}
    \end{equation}
     We obtain from Eq. (\ref{C-mav}) and Eq. (\ref{N-mav}) that if initial quantum states are maximally entangled and we make measurements in the Bell basis, then average concurrence and average negativity are equivalent. We simply obtain the average concurrence (average negativity) by taking the product of concurrences (negativities) of the initial states. Besides Eq. \eqref{C-av} shows that  measurement in non-maximally entangled basis during entanglement swapping degrades the average swapped entanglement.
    
    Now we extend entanglement swapping between two pairs of qubits to three pairs of qubits. We take three pairs of entangled qubits and make a measurement in the GHZ basis and analyze the outcome state. If the three pairs of qubits in Schmidt form are $\ket{\phi}_{AB} =\sqrt{\lambda_{0}}\ket{00}_{AB}+\sqrt{\lambda_{1}}\ket{11}_{AB})$, $\ket{\phi}_{CD} =\sqrt{\mu_{0}}\ket{00}_{CD}+\sqrt{\mu_{1}}\ket{11}_{CD}$ and $\ket{\phi}_{EF}\sqrt{\nu_{0}}\ket{00}_{EF}+\sqrt{\nu_{1}}\ket{00}_{EF}$ then the six-qubit system can be written as
     \begin{equation}\label{6-qubits}
       \ket{\Phi^{\prime}}=|\phi\rangle_{AB}\otimes\ket{\phi}_{CD}\otimes\ket{\phi}_{EF}.
    \end{equation}
   Let us make measurements on the $B$, $D$, and $F$ qubits. For this purpose, we can  define the triqubit GHZ basis as \cite{tsujimoto2018high}  
    
    \begin{equation}
        \begin{aligned}
            \left|G_{0,1}\right\rangle &=\frac{1}{\sqrt{2}}(|000\rangle \pm|111\rangle)\;, \\
            \left|G_{2,3}\right\rangle &=\frac{1}{\sqrt{2}}(|001\rangle \pm|110\rangle)\;, \\
            \left|G_{4,5}\right\rangle &=\frac{1}{\sqrt{2}}(|010\rangle \pm|101\rangle)\;, \\
            \left|G_{6,7}\right\rangle &=\frac{1}{\sqrt{2}}(|011\rangle \pm|100\rangle)\;.
        \end{aligned}
    \end{equation}
    Here the $-$ sign applies to states with odd indices. Now we can write Eq. (\ref{6-qubits}) in terms of GHZ basis as
    
     \begin{equation}\label{eq:6-qubitsfinal}
           \begin{aligned}
               \ket{\Phi^{\prime}}&=\frac{1}{\sqrt{2}}\left(\sqrt{\lambda _0\mu _0\nu _0}|000\rangle_{ACE} +\sqrt{\lambda _1\mu _1\nu _1}|111\rangle_{ACE} \right)|G_{0}\rangle\\
               &+\frac{1}{\sqrt{2}}\left(\sqrt{\lambda _0\mu _0\nu _0}|000\rangle_{ACE}
               -\sqrt{\lambda _1\mu _1\nu _1} |111\rangle_{ACE} \right)|G_{1}\rangle\\
               &+\frac{1}{\sqrt{2}}\left(\sqrt{\lambda _0\mu _0\nu _1}|001\rangle_{ACE}+\sqrt{\lambda _1\mu _1\nu _0}|110\rangle_{ACE} \right)|G_{2}\rangle\\
               &+\frac{1}{\sqrt{2}}\left((\sqrt{\lambda _0\mu _0\nu _1}|001\rangle_{ACE} -\sqrt{\lambda _1\mu _1\nu _0}|110\rangle_{ACE} \right)|G_{3}\rangle\\
               &+\frac{1}{\sqrt{2}}\left(\sqrt{\lambda _0\mu _1\nu _0}|010\rangle_{ACE} +\sqrt{\lambda _1\mu _0\nu _1}|101\rangle_{ACE}\right)|G_{4}\rangle\\
               &+\frac{1}{\sqrt{2}}\left(\sqrt{\lambda _0\mu _1\nu _0}|010\rangle_{ACE} -\sqrt{\lambda _1\mu _0\nu _1}|101\rangle_{ACE} \right)|G_{5}\rangle\\
               &+\frac{1}{\sqrt{2}}\left(\sqrt{\lambda _0\mu _1\nu _1}|011\rangle_{ACE} +\sqrt{\lambda _1\mu _0\nu _0}|100\rangle_{ACE} \right)|G_{6}\rangle\\
               &+\frac{1}{\sqrt{2}}\left(\sqrt{\lambda _0\mu _1\nu _1}|011\rangle_{ACE} -\sqrt{\lambda _1\mu _0\nu _0}|100\rangle_{ACE} \right)|G_{7}\rangle\;.
        \end{aligned}
    \end{equation}
    This equation shows that after measurements on $BDF$ qubits, we gain $ACE$ qubits in any one of the eight possible forms of entangled state. For example, if measurement gives us $|G_{0}\rangle$ then $ACE$ qubits have state
    \begin{equation}
        \frac{1}{\sqrt{2 p_{0}}}\left(\sqrt{\lambda _0\mu _0\nu _0}|000\rangle_{\text {ACE}}+\sqrt{\lambda _1\mu _1\nu _1}|111\rangle_{\text {ACE}}\right)\;.
    \end{equation}
    Here probability of getting $|G_{0}\rangle$ state is $p_{0}=\left(\lambda _0\mu _0\nu _0 +\lambda _1\mu _1\nu _1\right)/2$.
    
    Yu and Song \cite{yu2004free} showed that any good bipartite entanglement measure $M_{A-B}$ can be extended to multipartite systems by taking  bipartite partitions of them. So a tripartite entanglement quantifier can be defined as
    \begin{equation}
        M_{A B C}^{+}=\frac{1}{3}\left(M_{A-B C}+M_{B-A C}+M_{C-A B}\right) .
    \end{equation}
    But $M_{A B C}^{+}$ could be nonzero for pure biseparable states. It can be avoided by using the geometric mean:
    \begin{equation}
        M_{A B C}^{\times}=\left(M_{A-B C} M_{B-A C} M_{C-A B}\right)^{\frac{1}{3}}\;.
    \end{equation}
    Now by considering this bi-partition for the final triqubit entangled state, we can compute the swapped entanglement in the form of concurrence as  
     \begin{equation}
        \textbf{C}_{A C E}=\left(\textbf{C}_{A-C E} \textbf{C}_{C-A E} \textbf{C}_{E-A C}\right)^{\frac{1}{3}}\;,
    \end{equation}
    where $\textbf{C}_{A-C E}=\sqrt{2\left(1-\operatorname{Tr}\left(\rho_{A}^{2}\right)\right)}$ and similarly we can compute $ \textbf{C}_{C-A E}$, $\textbf{C}_{E-A C}$. Here, $\rho_{A}$ is the one-qubit reduced density matrix of the qubit $A$, obtained after tracing out the other qubits. The average concurrence for the final three qubits state now can be written as
    \begin{equation}\label{av_con}
        \textbf{C}_{A C E}^{av}=\textbf{C}_{AB} \textbf{C}_{CD} \textbf{C}_{EF}\;.
    \end{equation}
    It is again equal to the product of the concurrences of the initial three states.
    
    We can compute the the negativity of triqubit state $\rho_{ACE}$ as
    \begin{equation}
         \mathcal{N}(\rho_{ACE})=\left(\mathcal{N}_{A-C E} \mathcal{N}_{C-A E} \mathcal{N}_{E-A C}\right)^{\frac{1}{3}}\;,
    \end{equation}
    where $\mathcal{N}_{A-C E}=-2\sum_{i} \lambda_{i}\left(\rho_{ACE}^{T_A}\right)$, $\lambda_{i}\left(\rho_{ACE}^{T_A}\right)$ are the negative eigenvalues of $\rho_{ACE}^{T_A}$, partial transpose of $\rho_{ACE}$ with respect to subsystem $A$ is defined as $\left\langle i_{A}, j_{C E}\left|\rho_{ACE}^{T_A}\right| k_{A}, l_{CE}\right\rangle=\left\langle k_{A}, j_{CE}|\rho| i_{A}, l_{Ce}\right\rangle$ and similarly, we can define $\mathcal{N}_{C-A E}$, $\mathcal{N}_{E-A C}$. The average negativity of the final triqubit state can be written as
    \begin{equation}\label{n_con}
        \mathcal{N}_{A C E}^{av}=\mathcal{N}_{AB} \mathcal{N}_{CD} \mathcal{N}_{EF}\;,
    \end{equation}
    where $\mathcal{N}_{AB}, \mathcal{N}_{CD}$ and $\mathcal{N}_{EF}$ are the negativities of the initial three biqubit states.
    
    \subsection{Application of swapped entanglement}
    
    The final swapped entanglement between Alice and Danny's qubit is represented by Eq. \eqref{eq:swap2q} has wide range of application, however, we are interested in imposing it for teleportation of an unknown qubit state. Let, after the entanglement swapping Alice and Danny attain the state $\ket{\ddot{\Phi}_{+}}_{AD}$ that can also be written as 
    \begin{equation}\label{eq:ad_biqubit}
        \ket{\ddot{\Phi}_{+}}_{AD}=\left(a\ket{0}_{A}\ket{0}_{D}+b\ket{1}_{A}\ket{1}_{D}\right)\;,
    \end{equation}
    where $a=\alpha_{0}^{*}\sqrt{p_{0} p_{0}^{\prime}/p_{\tilde{\Phi}_{+}}}$, $b = \beta_{0}^{*}\sqrt{p_{1} p_{1}^{\prime}/p_{\tilde{\Phi}_{+}}}$ and $\lvert a\rvert^2+\lvert b\rvert^2=1$. If $a = b =\frac{1}{ \sqrt{2}}$ then Eq. \eqref{eq:ad_biqubit} is maximally entangled otherwise non-maximally entangled.
    Let Alice and Danny win a maximally entangled state after entanglement swapping and Alice wants to teleport an unknown quantum state to Danny. We denote the state that Alice wants to send as
    \begin{equation}\label{eq:unknownqubit}
        \ket{\chi}=\alpha|0\rangle+\beta|1\rangle\;,
    \end{equation}
    where $\lvert \alpha\rvert^2+\lvert \beta\rvert^2=1$. Now the state $\ket{\chi}$ can be teleported easily as described in Ref. \cite{nielsen2010quantum}. However, if $\lvert a\rvert\not =\frac{1}{\sqrt{2}}$ then Alice and Danny are not sharing a maximally entangled state and in this case, we use probabilistic teleportation to transmit $\ket{\chi}$. In such a situation, the receiver (Danny) cannot apply single-qubit unitary operations $I,X,iY,Z$ on his collapsed state $\frac{\alpha a|0\rangle + \beta b|1\rangle}{\sqrt{\left|\alpha a\right|^2+\left|\beta b\right|^2}}$ to obtain $\ket{\chi}$. Therefore, Danny has to prepare an ancilla qubit $\ket{0}_{Auxi}$ and applies $U_{0}$ unitary operation on the combined system as
    \begin{equation}\label{eq:prob-tele}
        U_{0}\left(\frac{\alpha a|0\rangle + \beta b|1\rangle}{\sqrt{\left|\alpha a\right|^2+\left|\beta b\right|^2}}\ket{0}_{Auxi}\right)\;,
    \end{equation}
    where 
    \begin{equation}
        U_{0}=\left(
                    \begin{array}{cccc}
                    \frac{b}{a} & \sqrt{1-\frac{b^2}{a^2}} & 0 & 0 \\
                    0 & 0 & 0 & -1 \\
                    0 & 0 & 1 & 0 \\
                    \sqrt{1-\frac{b^2}{a^2}} & -\frac{b}{a} & 0 & 0 \\
                    \end{array}
            \right)\nonumber\;.
    \end{equation}
    After implementation of unitary $U_{0}$, the expression \eqref{eq:prob-tele} attains the form
    \begin{equation}
        \frac{1}{\sqrt{\left|\alpha a\right|^2+\left|\beta b\right|^2}}\left(b\left(\alpha\ket{0}+\beta\ket{1}\right)\ket{0}+\alpha\sqrt{a^2-b^2}\ket{1}\ket{1}\right)\;.
    \end{equation}
    
    Now Danny makes a measurement on his ancilla (right most) qubit in the computational basis $\{\ket{0},\ket{1}\}$. If he gets $\ket{0}$ then his state collapses to $\alpha\ket{0}+\beta\ket{1}$, Danny further applies $I_{2\times 2}$ operation on the state obtained to reconstruct the desired state $\ket{\chi}$. If the measurement of ancilla gives $\ket{1}$ then protocol fails to teleport the required state due to its probabilistic nature. Similarly, we can also explain the teleportation of an unknown qubit for other states that appeared in Eq. \eqref{eq:swap2q}.  
   
    The measurement on Eq. \eqref{eq:6-qubitsfinal} gives us a three-qubit entangled state that can be any one of the eight three-qubit $GHZ$ states. The teleportation for the three-qubit $GHZ$ state has already been considered in Ref. \cite{xiong2016multiple, yan2010probabilistic}.
    
 \section{Noisy qubits and entanglement swapping}\label{nqaes}    
    We have used so far pure quantum systems. These quantum systems are isolated from external environments which comprise a variety of disorders and noises. In reality, quantum systems interact with the environment. One of the important types of noise is called depolarizing noise or white noise. This type of noise takes a quantum state and replaces it with a completely mixed state $\frac{1}{N}\mathbb{I}$, where $N$ is the dimension of the quantum system and $\mathbb{I}$ is identity matrix. Let us consider a biqubit noisy state that is prepared by mixing a pure state with white noise:
     \begin{equation}\label{noisyq}
           \begin{aligned}
               \rho_{\alpha}&=\alpha \rho_{AB}+(1-\alpha) \mathbb{I}_{2} \otimes \mathbb{I}_{2} /4\\
                            &=
                                \left(
                                    \begin{array}{cccc}
                                    \frac{1-\alpha }{4}+\alpha  p_0 & 0 & 0 & \alpha  \sqrt{p_0} \sqrt{p_1} \\
                                    0 & \frac{1-\alpha }{4} & 0 & 0 \\
                                    0 & 0 & \frac{1-\alpha }{4} & 0 \\
                                    \alpha  \sqrt{p_0} \sqrt{p_1} & 0 & 0 & \frac{1-\alpha }{4}+\alpha  p_1 \\
                                    \end{array}
                                \right)\;, 
             \end{aligned}
    \end{equation}
    where $\rho_{AB}=\ket{\phi}_{AB}\bra{\phi}$ is the density operator of biqubit system $AB$ described in Eq. \eqref{eq5}, $\mathbb{I}_{2}$ identity matrix and parameter $\alpha$ called visibility of system $AB$. If we take $p_{0}=p_{1}=1/2$, the Eq. \eqref{noisyq} becomes an \textit{isotropic state} \cite{horodecki1999reduction} with maximally entangled $\rho_{AB}$. The isotropic states are invariant under all transformations of the form $U\otimes U^{*}$, where the asterisk denotes complex conjugation in a certain basis. 
    
\begin{figure}
        \centering
        \includegraphics[width=0.7\textwidth]{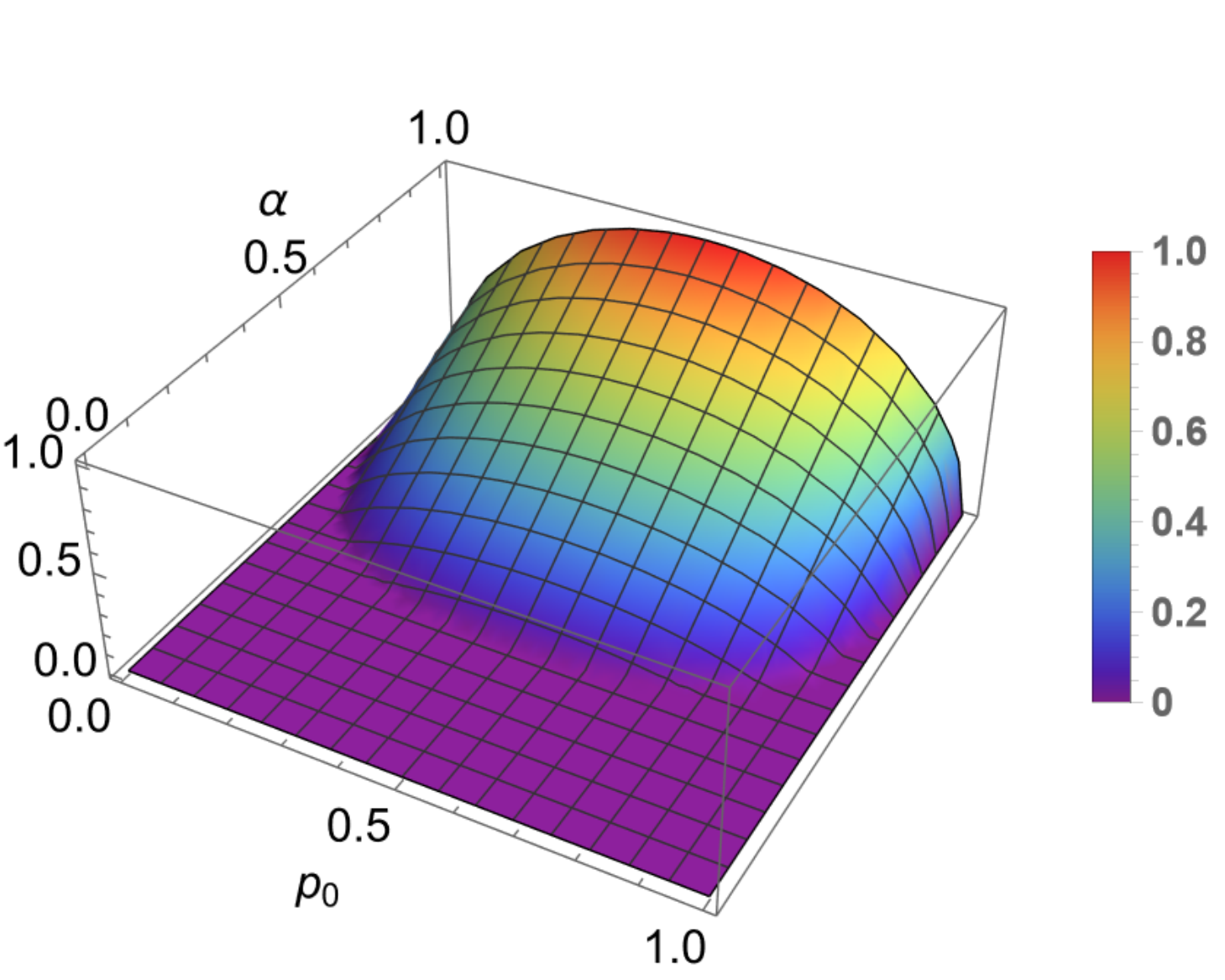}
        \hfill
        \caption{This plot represents entangled states present in the mixed state $ \rho_{\alpha}$. All entangled states have $f(\alpha,p_{0})= \|\tilde{\mathcal{T}}\|_{KF}-1>0$.}
        \label{fig:2qubit-noisy-state}
    \end{figure}

    We can also represent a biqubit noisy system in the Bloch form as \cite{zangi2021combo}
    \begin{equation}
        \begin{aligned}
        \rho=& \frac{1}{4} \mathbb{I}_{2} \otimes \mathbb{I}_{2}+\sum_{\mu=1}^{3} r_{\mu} \frac{\sigma_{\mu}}{\sqrt{2}} \otimes \frac{\mathbb{I}_{2}}{2}+\sum_{\nu=1}^{3} s_{\nu} \frac{\mathbb{I}_{2}}{2} \otimes \frac{\sigma_{\nu}}{\sqrt{2}} 
        &+\sum_{\mu=1}^{3} \sum_{\nu=1}^{3} t_{\mu \nu} \frac{\sigma_{\mu}}{\sqrt{2}} \otimes \frac{\sigma_{\nu}}{\sqrt{2}}\;,
        \end{aligned}
    \end{equation}
    where $\sigma$ represents Pauli matrices, $\mathbb{I}$ is identity matrix, $r_{\mu}=\operatorname{Tr}\left(\rho\frac{\sigma_{\mu}}{\sqrt{2}} \otimes \frac{\mathbb{I}_{2}}{2}\right)$ and $s_{\nu}=\operatorname{Tr}\left(\rho \frac{\mathbb{I}_{2}}{2} \otimes \frac{\sigma_{\nu}}{\sqrt{2}}\right)$ are Bloch vectors of given two qubits and $t_{\mu \nu}=\operatorname{Tr}\left(\rho \frac{\sigma_{\mu}}{\sqrt{2}} \otimes \frac{\sigma_{\nu}}{\sqrt{2}}\right)$ called a correlation tensor. We can construct Bloch matrix from $\Vec{r}$, $\Vec{s}$ and $3\times 3$ dimensional correlation matrix $T$ as 
    \begin{equation}\label{Bloch_formj}
         \tilde{\mathcal{T}}=
         \left(
            \begin{array}{cc}
                c & \Vec{s} \\
                \Vec{r} & T \\
            \end{array}
        \right)\;,
    \end{equation}
    where $c$ is a scalar number. The Bloch matrix form of Eq. \eqref{noisyq} contains $c =\alpha  \sqrt{p_0} \sqrt{p_1}$, $\Vec{r}=\Vec{s}=0$ and correlation matrix
    \begin{equation}\label{Bloch_form}
    \setlength{\arraycolsep}{2pt}
  \renewcommand{\arraystretch}{0.8}
         T=
         \left(\scriptsize
            \begin{array}{ccc}
             -\alpha  \sqrt{p_0} \sqrt{p_1} & 0 & 0 \\
             0 & \frac{\alpha -1}{4}+\frac{1}{2} \left(\frac{1-\alpha }{4}+\alpha  p_0\right)+\frac{1}{2} \left(\frac{1-\alpha }{4}+\alpha  p_1\right) & \frac{1-\alpha }{8}+\frac{\alpha -1}{8}+\frac{1}{2} \left(\frac{1-\alpha }{4}+\alpha  p_0\right)+\frac{1}{2} \left(\frac{\alpha -1}{4}-\alpha  p_1\right) \\
             0 & \frac{1-\alpha }{8}+\frac{\alpha -1}{8}+\frac{1}{2} \left(\frac{1-\alpha }{4}+\alpha  p_0\right)+\frac{1}{2} \left(\frac{\alpha -1}{4}-\alpha  p_1\right) & \frac{1-\alpha }{4}+\frac{1}{2} \left(\frac{1-\alpha }{4}+\alpha  p_0\right)+\frac{1}{2} \left(\frac{1-\alpha }{4}+\alpha  p_1\right) \\
            \end{array}
         \right)\;. 
    \end{equation}
    As the coherence vectors $\Vec{r}$, $\Vec{s}$ of the subsystems $A$ and $B$ have zero magnitudes that means the state is maximally mixed. According to combo separability criteria \cite{zangi2021combo} if $f(\alpha,p_{0})= \|\tilde{\mathcal{T}}_{\alpha}\|_{KF}-1>0$ then state $ \rho_{\alpha}$ is an entangled state.
    We plotted $f(\alpha,p_{0})$ in Fig. (\ref{fig:2qubit-noisy-state}) which represents entanglement of the mixed state $ \rho_{\alpha}$.
    It is clear from the figure that $\rho_{\alpha}$ remains entangled when $0<p_{0}<1$ and the minimum value of $\alpha$ is $1/3$. This state becomes maximally entangled  when $p_{0}=0.5$ and $\alpha$ approaches 1.
    
\begin{figure}
            \centering
            \begin{subfigure}[t]{0.49\textwidth}
            \centering
            \includegraphics[width=\textwidth]{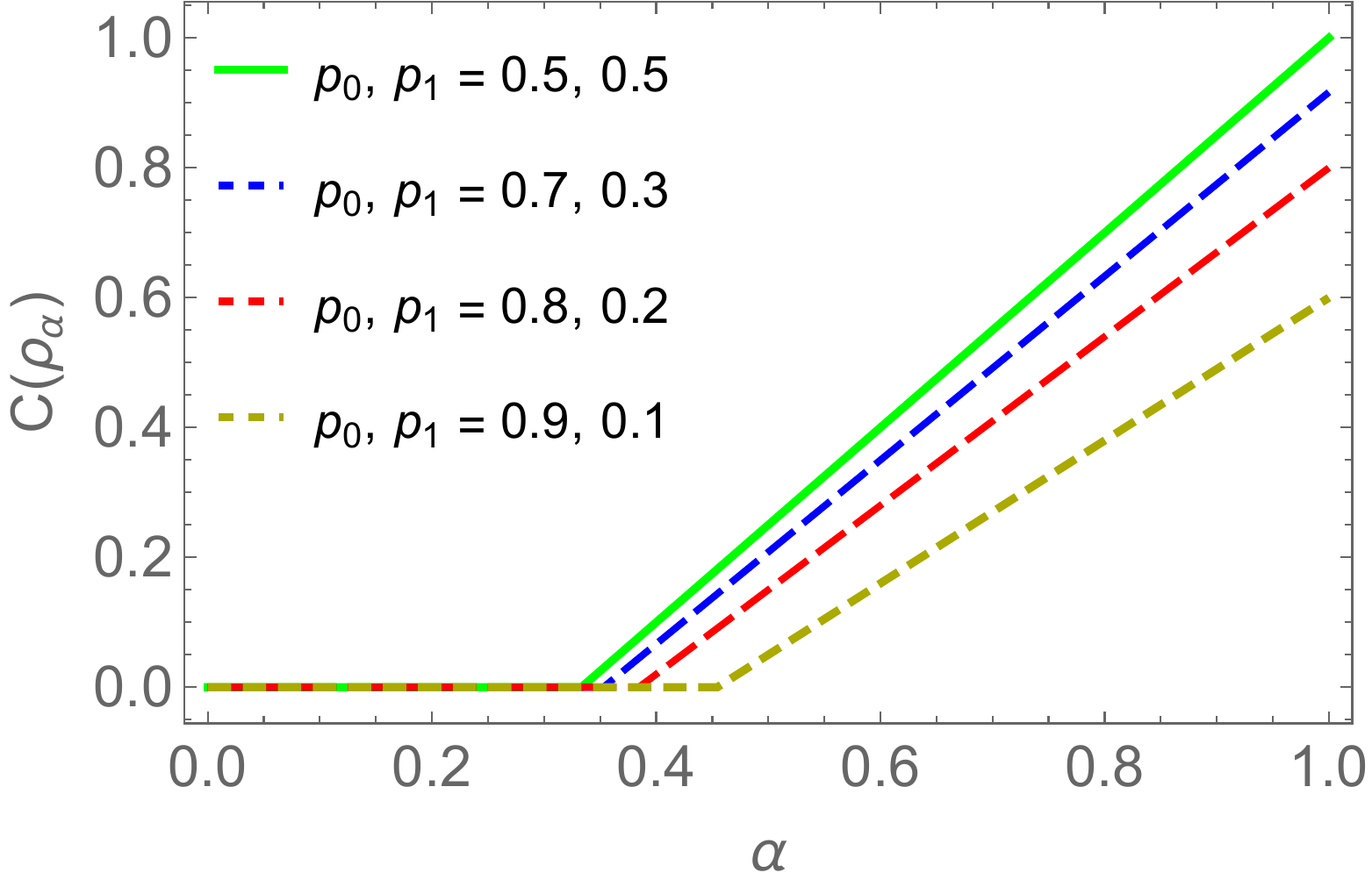}
            \caption{}
            \label{fig:a-concurrence}
            \end{subfigure}
            \hfill
            \begin{subfigure}[t]{0.49\textwidth}
            \centering
            \includegraphics[width=\textwidth]{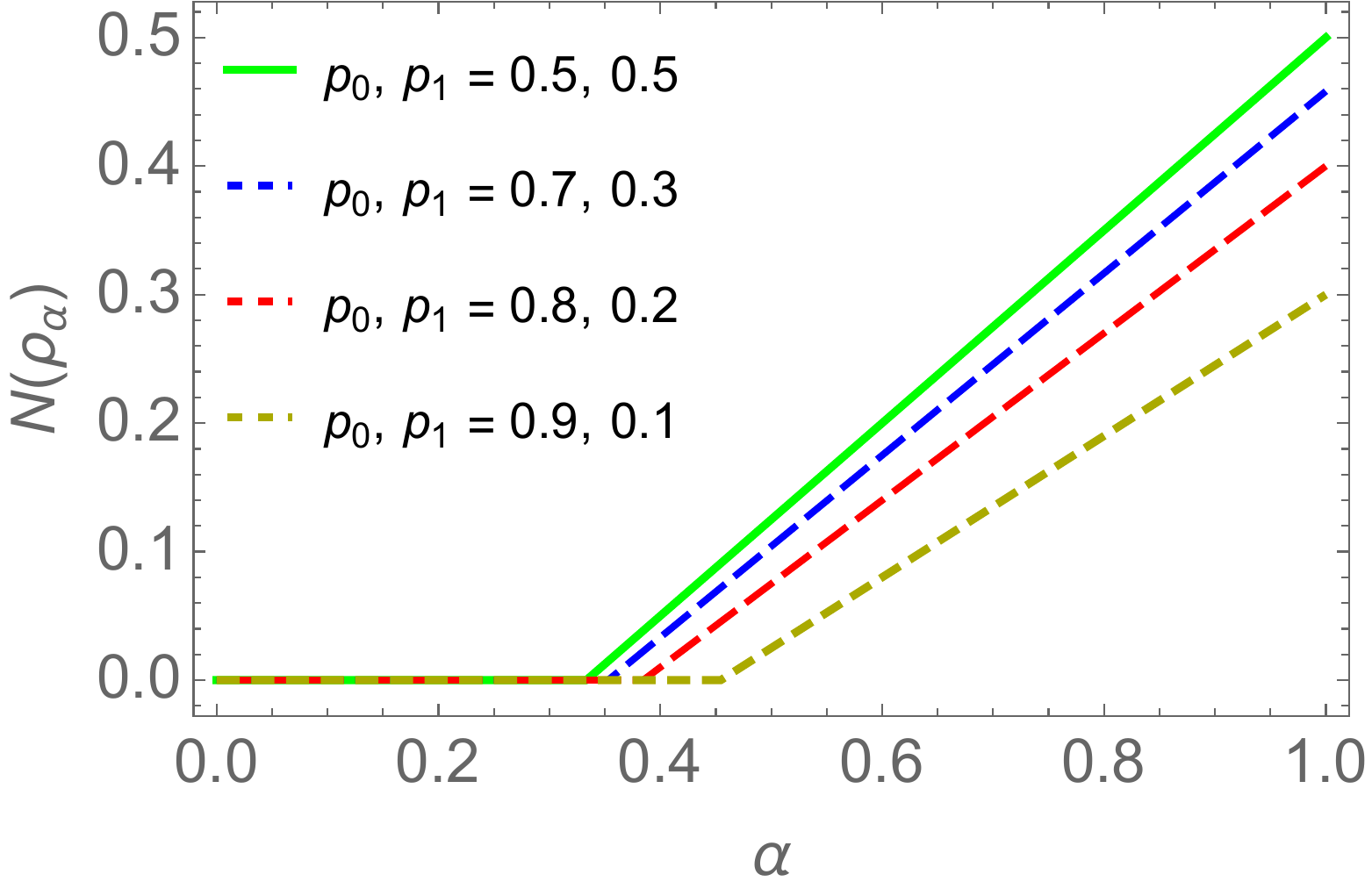}
            \caption{}
            \label{fig:b-negativity}
            \end{subfigure}
            \hfill
     
            \caption{Concurrence and negativity comparison of biqubit noisy states. Fig. (a) represents the concurrence of biqubit state against state visibility parameter $\alpha$ and similar Fig. (b) represents the negativity of biqubit state.}
            \label{fig:biqubit-NC}
    \end{figure}

    As $\rho_{\alpha}$ is a X-form mixed-state with nonzero entries only along the diagonal and anti-diagonal so its concurrence is given by \cite{quesada2012quantum}
    \begin{equation}\label{X1-form-conc}
        \begin{aligned}
            \textbf{C}({\rho_\alpha})&=\max \left[0,2\left(\left|\rho_{\alpha}^{(14)}\right|-\sqrt{\rho_{\alpha}^{(22)} \rho_{\alpha}^{(33)}}\right),2\left(\left|\rho_{\alpha}^{(23)}\right|-\sqrt{\rho_{\alpha}^{(11)} \rho_{\alpha}^{(44)}}\right)\right]\\
            &= \max \left[0,2\left(\alpha\sqrt{p_0 p_1}-\frac{1-\alpha}{4}\right)\right]\;.
        \end{aligned}
    \end{equation}
    This relation also gives a lower bound for the probability that keeps the $\rho_{\alpha}$ entangled as
    \begin{equation}
      \alpha>\frac{1}{\left(1+4 \sqrt{p_{0}p_{1}}\right)}\;.
    \end{equation}
    If $\rho_{AB}$ is a maximally entangled state then $p_{0}=p_{1}=\frac{1}{2}$, in this case the state remains entangled for $\alpha>\frac{1}{3}$ that we can also observe from Fig (\ref{fig:a-concurrence}).
    
    The negativity of $X-$form state $\rho_{\alpha}$ can be computed  as
    \begin{equation}\label{neg-x}
        \mathcal{N}(\rho_{X})=-2\min \left\{0,r_{+}-\sqrt{r_{-}^{2}+\left(\rho^{(14)}\right)^2},u_{+}-\sqrt{u_{-}^{2}+\left(\rho^{(23)}\right)^2}\right\}\;,
    \end{equation}
    where $u_{\pm}=\left(\rho^{(11)}\pm\rho^{(44)}\right)/2$, $r_{\pm}=\left(\rho^{(22)}\pm\rho^{(33)}\right)/2$.
    The Eq. \eqref{neg-x} can be reduced to 
    \begin{equation}\label{red-neg-x}
        \mathcal{N}(\rho_{\alpha})=-2\min \left\{0,\frac{1}{4} \left(1-\alpha -4 \alpha \sqrt{ p_0 p_1}\right) \right\}\;,
    \end{equation}
    because $\rho^(23)=0$ and $u_{+}-u_{-}=\rho^{(44)}$ is a positive number. For a maximally entangled state, the Eq. \eqref{red-neg-x} also gives $\alpha>\frac{1}{3}$ for entanglement retain and can be observed from Fig. (\ref{fig:b-negativity}).
    It is clear from Fig. (\ref{fig:biqubit-NC}) that the concurrence and negativity produce similar results in the case of biqubit noisy state. 
    
         \begin{figure}
            \centering
            \begin{subfigure}[t]{0.49\textwidth}
            \centering
            \includegraphics[width=\textwidth]{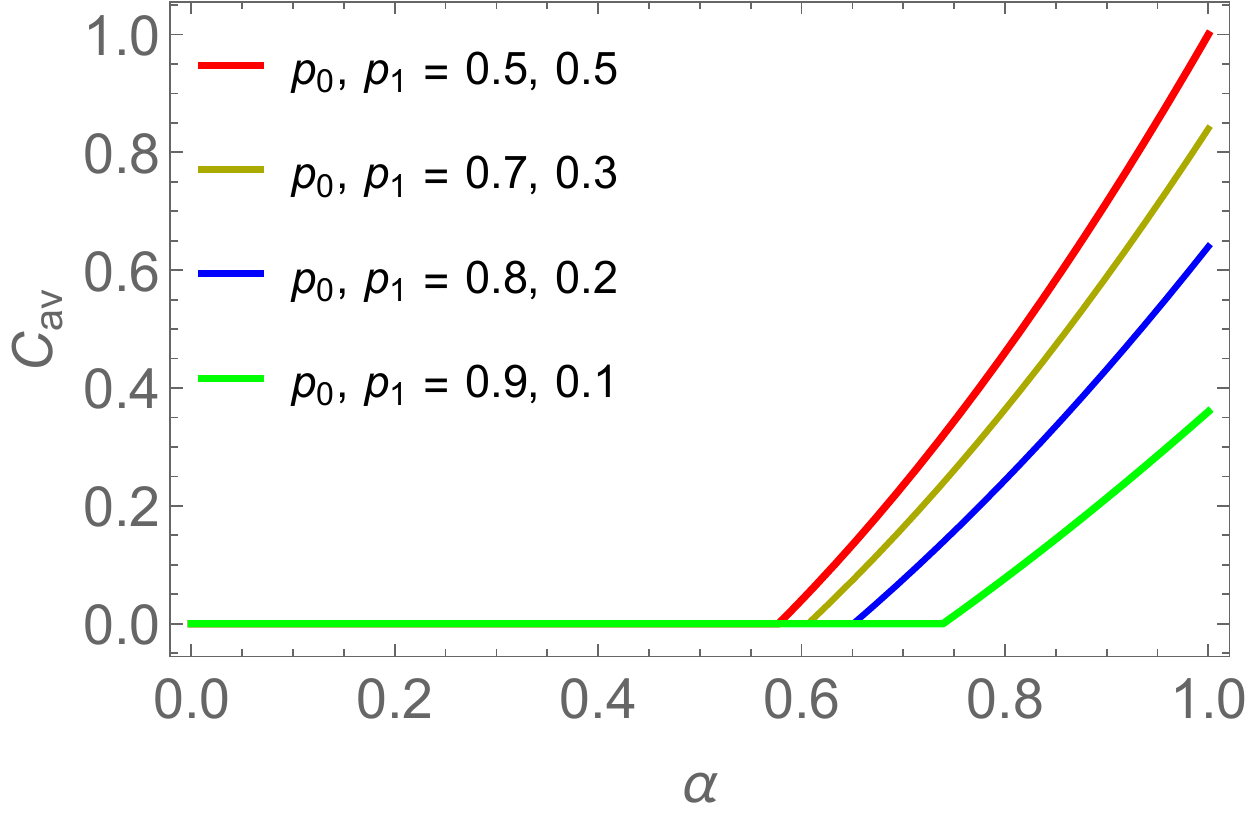}
            \caption{}
            \label{fig:cav-plot}
            \end{subfigure}
            \hfill
            \begin{subfigure}[t]{0.49\textwidth}
            \centering
            \includegraphics[width=\textwidth]{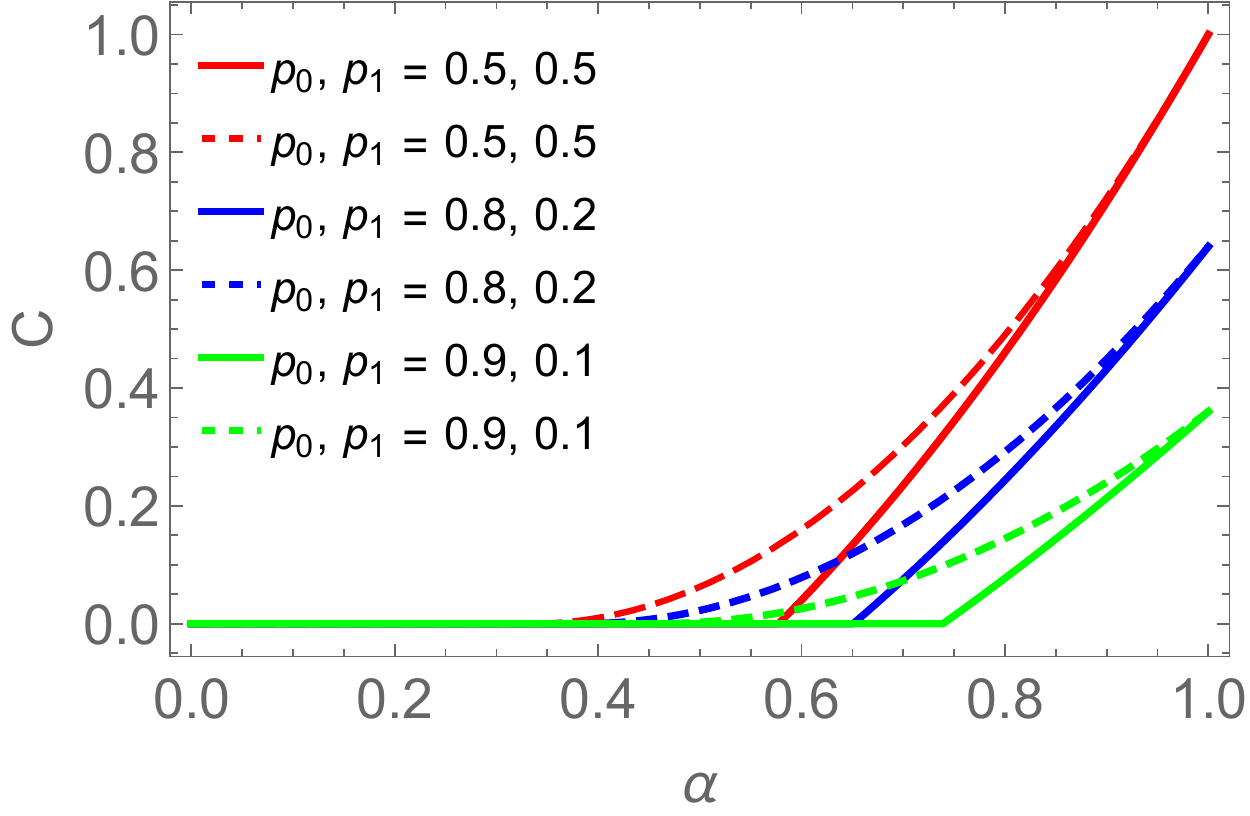}
            \caption{}
            \label{fig:cav-cpro-plot}
            \end{subfigure}
            \hfill
     
            \caption{Average concurrence of final states. Fig. (a) shows four different types of plots of average concurrence for the final states depending upon the different values of $p_{0}$ and $p_{1}$ against $\alpha$. Fig. (b) is comparison of the average concurrence of final states (solid lines) with the product of concurrences of input states (dashed lines).}
            \label{fig:ave-noisy-conc}
    \end{figure}
     
    Now we want to explore entanglement swapping between two states of the form given in Eq. \eqref{noisyq}. For simplicity, we assume that the states are similar and we make standard Bell measurements in order to accomplish the entanglement swapping. Therefore our four-qubit noisy state in terms of Bell basis is
    \begin{equation}\label{4noisyqBell}
        \begin{aligned}
            \rho_{ABCD}=&\frac{\alpha ^2}{2}\left(p_0 |0 0\rangle_{AD} \pm p_1 |1 1\rangle_{AD}\right)\left(p_0 \langle 0 0|\pm p_1 \langle 1 1|\right)|\Phi _{\pm }\rangle _{\text{BC}}\langle \Phi _{\pm }| \\
            &+\frac{\alpha ^2 p_0 p_1}{2}\left(|01\rangle_{AD} \pm |1 0\rangle_{AD}\right)\left(\langle 0 1|\pm \langle 1 0|\right)|\Psi _{\pm }\rangle _{\text{BC}}\langle \Psi _{\pm }|\\
            &+\mathbf{d}_{AD}\left(|\Phi _{\pm }\rangle _{\text{BC}}\langle \Phi _{\pm }|+|\Psi _{\pm }\rangle _{\text{BC}}\langle \Psi _{\pm }|\right)\;,
        \end{aligned}
    \end{equation}
    where 
    \begin{multline}
        \mathbf{d}_{AD}=\frac{\alpha\left(1-\alpha\right)}{8}\left(\left(p_{0}\ket{0}_{A}\bra{0}+p_{1}\ket{1}_{A}\bra{1}\right)\otimes\mathbb{I}_{D}+\mathbb{I}_{A}\otimes\left(p_{0}\ket{0}_{D}\bra{0}+p_{1}\ket{1}_{D}\bra{1}\right)\right)\\
        +\frac{(1-\alpha)^2}{16}\mathbb{I}_{AD}s\;,
    \end{multline}
    and $\mathbb{I}$ is an identitiy matrix. The measurement of the qubits $B$ and $C$ will give us one of the following four Bell states
    \begin{equation}\label{Bell-basis}
        \begin{aligned}
            \left|\Phi_{\pm}\right\rangle_{BC} &=\frac{1}{\sqrt{2}}\left(|00\rangle_{BC} \pm|11\rangle_{BC}\right)\;, \\
            \left|\Psi_{\pm}\right\rangle_{BC} &=\frac{1}{\sqrt{2}}\left(|01\rangle_{BC} \pm|10\rangle_{BC}\right)\;.
        \end{aligned}
    \end{equation}
    If we obtain $\left|\Phi_{\pm}\right\rangle_{BC}$ then the qubits $A$ and $D$ become entangled in the state
    \begin{equation}\label{r_AD_phi}
    \rho_{AD}^{\Phi_{\pm}}=\frac{1}{P_{\Phi}}\left(\frac{\alpha^2}{2}\left(p_{0}|00\rangle_{AD} \pm p_{1}|11\rangle_{AD}\right)\left(p_{0}\langle00|\pm p_{1}\langle11|\right)+\mathbf{d}_{AD}\right)\;,
    \end{equation}
    where $P_{\Phi}=\frac{\alpha^2}{2}\left(p_{0}^{2}+p_{1}^{2}\right)+\frac{1-\alpha^2}{4}$ is the probability of $\ket{\Phi_{+}}$ and $\ket{\Phi_{-}}$. If measurement gives us $\left|\Psi_{\pm}\right\rangle_{BC}$ then the qubits $A$ and $D$ make entangled state
    \begin{equation}\label{r_AD_psi}
        \rho_{AD}^{\Psi_{\pm}}= \frac{1}{P_{\Psi}}\left(\frac{\alpha^2 p_{0}p_{1}}{2}\left(|01\rangle_{AD} \pm|10\rangle_{AD}\right)
        \left(\langle 01| \pm\langle 10|\right)+\mathbf{d}_{AD}\right)\;.
    \end{equation}
    Here, $P_{\Psi}=\alpha^2 p_{0}p_{1}+\frac{1-\alpha^2}{4}$ is the probability of $\ket{\Psi_{+}}$ and $\ket{\Psi_{-}}$.
    
   In order to evaluate the transferred entanglement between qubits $A$ and $D$, we first compute the concurrence of all four types of density matrix $\rho_{AD}$. As all density matrices of qubits $A$ and $D$  in Eq. \eqref{r_AD_phi} and Eq. \eqref{r_AD_psi} have X-form state form so, their concurrence can easily be computed by using Eq. \eqref{X1-form-conc}. The state $\rho_{AD}^{\Phi_{+}}$ and $\rho_{AD}^{\Phi_{-}}$ have the same amount of concurrence and is given by
    \begin{equation}\label{noisy-conc-phi}
        \textbf{C}\left( \rho_{AD}^{\Phi_{\pm}}\right)=\frac{1}{P_{\phi }}\max \left(0,\alpha ^2 p_0 p_1-\frac{1}{8} \left(1-\alpha ^2\right)\right)\;.
    \end{equation}
    and similarly, the concurrence of $\rho_{AD}^{\Psi_{\pm}}$ is 
    \begin{equation}\label{noisy-conc-psi}
        \begin{aligned}
            \textbf{C}\left( \rho_{AD}^{\Psi_{\pm}}\right)=\frac{1}{P_{\psi }}\max \left(0,\alpha ^2 p_0 p_1-\frac{1}{8} (1-\alpha ) \sqrt{1+2 \alpha-3 \alpha ^2 +16 \alpha ^2 p_0 p_1}\right)\;.
        \end{aligned}
    \end{equation}
    Now the average of teleported entanglement in terms of concurrence can be computed as
    \begin{equation}\label{ave-conc}
        \textbf{C}_{a v}=2 P_{\Phi} \textbf{C}\left(\rho_{A D}^{\Phi_{+}}\right)+2 P_{\Psi} \textbf{C}\left(\rho_{A D}^{\Psi_{+}}\right)\;.
    \end{equation}
   This average concurrence of the final states has been plotted in Fig. (\ref{fig:cav-plot}). Besides Fig. (\ref{fig:cav-cpro-plot}) put forward that for mixed states, the product of the concurrences of the initial states (dashed line plots) is an upper bound to the average concurrence of the finally swapped entanglement (solid line plots).
    
\begin{figure}
            \centering
            \begin{subfigure}[t]{0.49\textwidth}
                \centering
                \includegraphics[width=\textwidth]{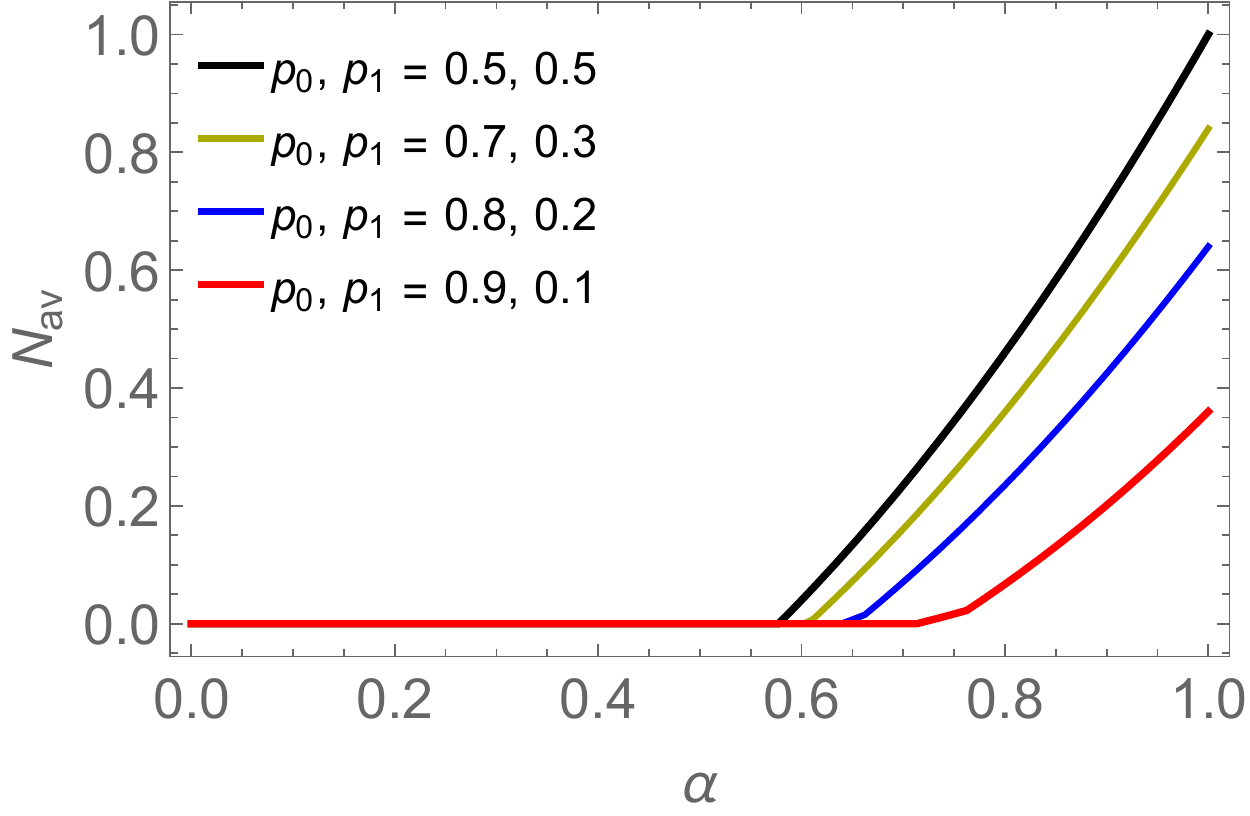}
                \caption{}
                \label{fig:Nav-plot}
            \end{subfigure}
            \hfill
            \begin{subfigure}[t]{0.49\textwidth}
                \centering
                \includegraphics[width=\textwidth]{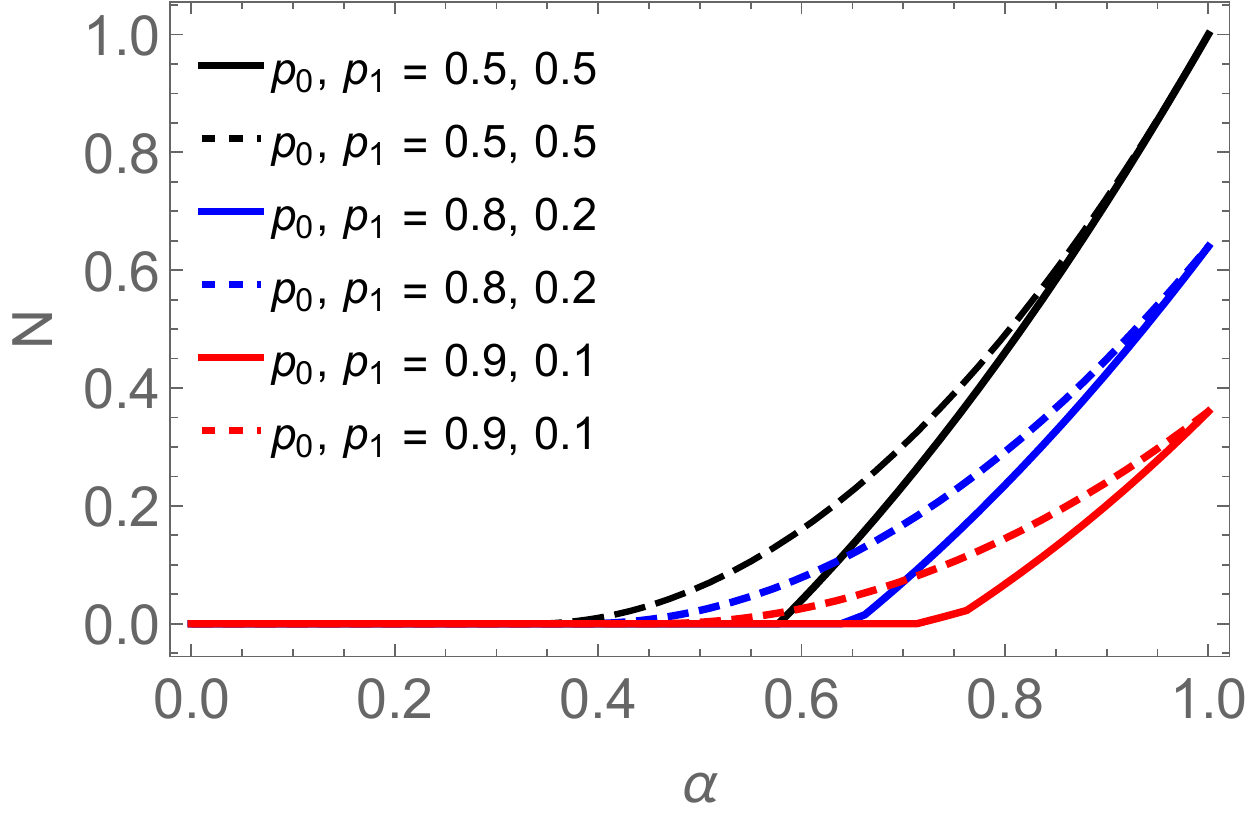}
                \caption{}
                \label{fig:Nav-Npr-plot}
            \end{subfigure}
            \hfill
     
            \caption{Average swapped entanglement in terms of negativity. Fig. (a) Average negativity of final states. Fig. (b) shows the comparison of the average negativity of final states (solid lines) with the product of negativities of initial states (dashed lines).}
            \label{fig:neg}
    \end{figure}

\begin{figure}
        \centering
        \includegraphics[width=0.7\textwidth]{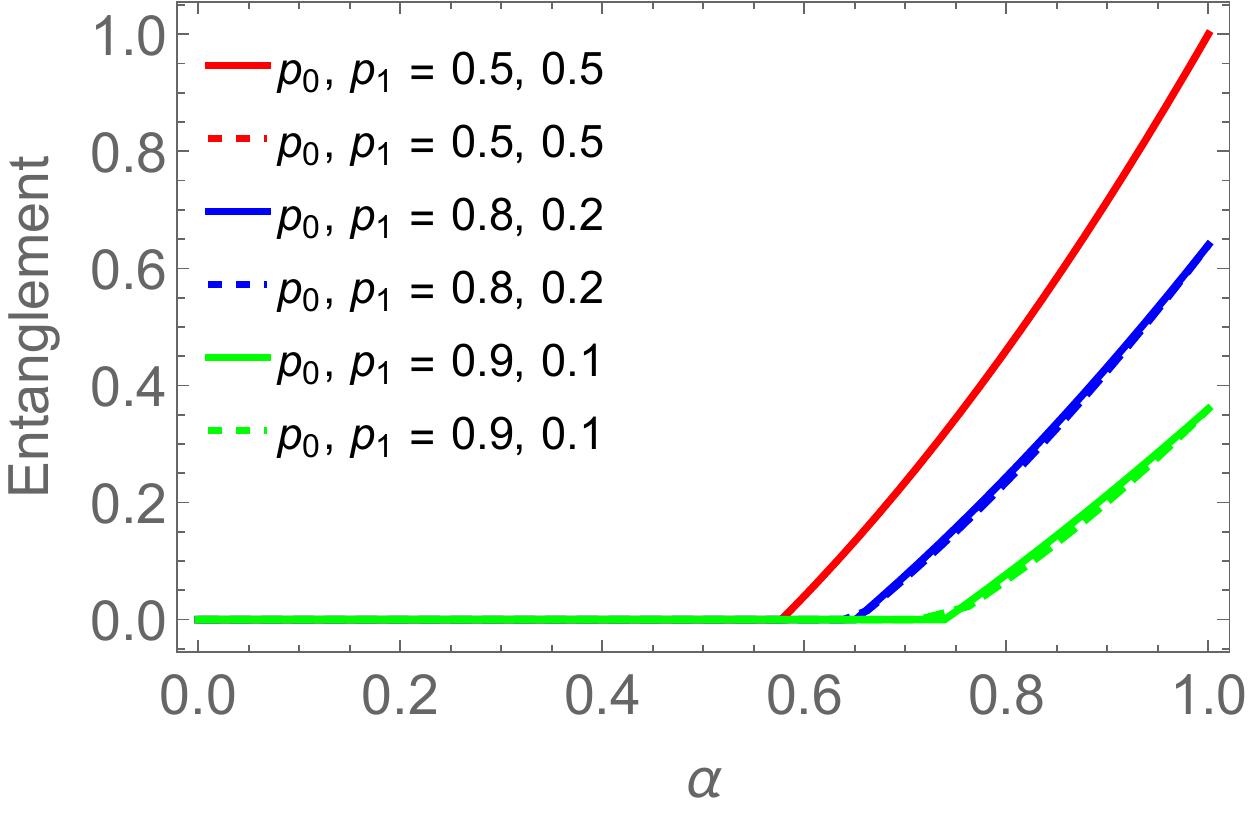}
        \hfill
        \caption{Comparison of entanglement quantifiers for swapped entanglement. Solid lines represent the average concurrence and the dashed lines represent the average negativity of the final states}.
        \label{fig:ave-conc-neg}
    \end{figure}

    We can also evaluate the transferred entanglement between qubits $A$ and $D$ in terms of negativity. As all density matrices of qubits $A$ and $D$  in Eq. \eqref{r_AD_phi} and Eq. \eqref{r_AD_psi} have X-form state, hence, their negativity can easily be computed by Eq. \eqref{neg-x}. The state $\rho_{AD}^{\Phi_{+}}$ and $\rho_{AD}^{\Phi_{-}}$ have the same amount of negativity and take the form
     \begin{equation}\label{noisy-neg-phi}
        \mathcal{N}\left( \rho_{AD}^{\Phi_{\pm}}\right)=-\frac{2}{P_{\phi }}\min \left(0,\frac{1}{16} \left(1-\alpha ^2\right)-\frac{1}{2}\alpha ^2 p_0 p_1\right)\;.
    \end{equation}
    and 
    \begin{equation}\label{noisy-neg-psi}
        \begin{aligned}
            \mathcal{N}\left( \rho_{AD}^{\Psi_{\pm}}\right)
            =-\frac{2}{P_{\psi }}\min \left(0,\frac{1-\alpha ^2}{16} -\sqrt{\frac{1}{4} \alpha ^4 p_0^2 p_1^2+\frac{1}{64} (1-\alpha )^2 \alpha ^2 \left(p_0-p_1\right){}^2}\right)\;.
        \end{aligned}
    \end{equation}
    The average of swapped entanglement computed by negativity can be given as
    \begin{equation}\label{ave-neg}
        \mathcal{N}_{a v}=2 P_{\Phi}\mathcal{N}\left( \rho_{AD}^{\Phi_{\pm}}\right) +2 P_{\Psi}\mathcal{N}\left( \rho_{AD}^{\Psi_{\pm}}\right)\;.
    \end{equation}
    The average of swapped entanglement in terms of negativity has been plotted in Fig. (\ref{fig:neg}) and here, Fig. (\ref{fig:Nav-Npr-plot}) shows that the product of the negativities of the initial states (dashed line plots) provides an upper bound to the average negativity of the final states (solid line plots). Moreover, Fig. (\ref{fig:ave-conc-neg}) represents the comparison of average concurrence and average negativity of final states. This plot shows that when $p_{0}=p_{1}=0.5$ which correspond to maximally entangled input states then concurrence and negativity overlap but for other cases, concurrence provides an upper bound to the negativity for qubit systems.

   \subsection{Teleportation using a noisy quantum channel}
    
    The four Bell states for qubits $B$ and $C$ are given in Eq. \eqref{Bell-basis}. By using the similar Bell states for qubits $A$ and $D$, we can define the projectors $P^{\Phi_{\pm}}=\ket{\Phi_{\pm}}_{AD}\bra{\Phi_{\pm}}$ and $P^{\Psi_{\pm}}=\ket{\Psi_{\pm}}_{AD}\bra{\Psi_{\pm}}$ associated with the measurements that Alice performs in the execution of the teleportation protocol. Eqs. (\ref{r_AD_phi}, \ref{r_AD_psi}) represent the density matrices of four possible outcomes after entanglement swapping of noisy entangled states. We can use these density matrices as a channel for the teleportation of an unknown qubit given in Eq. \eqref{eq:unknownqubit} from Alice to Danny. Let the teleportation channel is $\rho_{AD}^{\Phi_{+}}$ and the density matrix of the qubit to be teleported is given by $\rho_{tq}=\ket{\chi}\bra{\chi}$, where the subscripts {\it tq} means ``teleportation qubit". The initial three qubits state is given by
    \begin{equation}
        \varrho_{1}=\rho_{tq} \otimes \rho_{AD}^{\Phi_{+}} .
    \end{equation}
    The first two qubits (i.e $\rho_{tq}$ and first qubit of $\rho_{AD}^{\Phi_{+}}$) of $ \varrho_{1}$ are possessed by Alice and the third qubit is occupied by Danny. Alice makes a projective measurement on her two qubits. After this measurement, we attain the post-measurement state
    \begin{equation}
        \tilde{\varrho_{1}}=\frac{P^{\Phi_{+}} \varrho_{1} P^{\Phi_{+}}}{\tilde{P_{1}}}\;,
    \end{equation}
    where $\tilde{P_{1}}=\operatorname{Tr}\left[P^{\Phi_{+}} \varrho_{1}\right]$ is the probability of occurrence of state $\tilde{\varrho_{1}}$ and Tr represents trace operation. Alice then communicates her outcomes with Danny via the classical channel. The qubit possessed by Danny has the form $\tilde{\varrho}_{D1}=\operatorname{Tr}_{12}\left[\tilde{\varrho_{1}}\right]$, where $\operatorname{Tr}_{12}$ means the partial trace of qubits 1 and 2. Due to the noisy teleportation channel, Danny has to follow a probabilistic teleportation technique to find teleported qubit $\rho_{tq}$. He prepares an auxiliary qubit $\rho_{Auxi}=\ket{0}\bra{0}$ and applies a suitable unitary operator $U_{i}$ on two qubits system as
    \begin{equation}
       U_{i}\left(\tilde{\varrho}_{D1}\otimes\rho_{Auxi}\right)U_{i}^\dagger\;.
    \end{equation}
    Then a measurement on Danny's auxiliary qubit in the basis $\left\{\ket{0}\bra{0},\ket{0}\bra{1},\ket{1}\bra{0},\ket{1}\bra{1}\right\}$ is done. If $\ket{0}\bra{0}$ occurs, we obtain qubit $\tilde{\varrho}^{\prime}_{D1}$ with some probability $P^{\prime}_{1}$ otherwise the teleportation fails.
    
    The protocol ends with Danny apply a unitary operation $u$ on his qubit final state as
    \begin{equation}
       \rho^{\prime}_{tq}= u\frac{\tilde{\varrho}^{\prime}_{D1}}{P^{\prime}_{1}}u^{\dagger}\;.
    \end{equation}
    The unitary operator $u$ is one of the Pauli operators $\left\{ \mathbb{I},\sigma_{x},\sigma_{y},\sigma_{z} \right\}$, and its choice depends not only on the measurement result of Alice but also on the quantum channel shared between Alice and Danny in the teleportation protocol.
    
    Now we can check the efficiency of the protocol by using fidelity \cite{uhlmann1976transition}. Since the input state $\rho_{tq}=\ket{\chi}\bra{\chi}$ is pure, the fidelity can be written as
      \begin{equation}
        F=\operatorname{Tr}\left[\rho_{tq}\rho^{\prime}_{tq}\right]=\bra{\chi}\rho^{\prime}_{tq}\ket{\chi}\;.
    \end{equation}
    The fidelity ranges from 0 to 1 and its maximal value occurs whenever the Danny's qubit final state $\rho^{\prime}_{tq}$ is equal to input qubit $\rho_{tq}$ and it is 0 when the two states are orthogonal.
    
   \section{Concluding discussion}\label{cd}
   
   We have studied an entanglement swapping protocol, where Alice and Bob share a generalized Bell pair $(A, B)$ whereas, Cara and Danny share another generalized Bell pair $(C, D)$. When Bob and Cara, who are situated in the same laboratory perform some measurements on the pair $(B, C)$ then initially unentangled qubits $(A, D)$ obtain entanglement. Alice and Danny know about the entanglement of their qubits after getting information about the qubits of Bob and Cara via classical communication channel.  
   
   In the case of two couples of pure qubits, the finally entangled couple can have one of the four possible entangled states. However, if we considered three couples of entangled qubits then entanglement swapping gives us a three-qubit entangled state that can be any one of the eight possible forms of GHZ quantum states. 
   
   The significant achievements of this study can be summarized as, if initial quantum states are maximally entangled and we make measurements in the Bell basis, then average concurrence and average negativity of final states give similar results. We simply obtain the average swapped entanglement among final quantum states by taking the product of entanglement of the initially  maximally entangled states. The measurement in non-maximally entangled basis during entanglement swapping degrades the swapped entanglement. The product of the entanglement of the mixed states provides an upper bound to the average swapped entanglement of final states. The entanglement quantifier concurrence provides an upper bound to the negativity. We also use the final output state as a channel for the teleportation of an unknown qubit from Alice to Danny. The teleportation with a pure biquibit Bell state is obvious, but we explored the probabilistic teleportation of an unknown qubit not only with non-maximally entangled channel but also with the noisy channel that we obtain after entanglement swapping.

   \section*{Acknowledgements}
   Financial support from the Postdoctoral training funds, grant nos. C615300501. S. M. Zangi is extremely grateful for the help and support of Bo Zheng. Chitra Shukla thanks to Shenzhen Science Technology and Innovation Commission: Grant number: GJHZ2020081109520301, and NSFC-Guangdong joint Fund: Grant number: U1801661.

    \bibliographystyle{IEEEtran}
    \bibliography{main.bib}

% Generated by IEEEtran.bst, version: 1.14 (2015/08/26)
\begin{thebibliography}{10}
\providecommand{\url}[1]{#1}
\csname url@samestyle\endcsname
\providecommand{\newblock}{\relax}
\providecommand{\bibinfo}[2]{#2}
\providecommand{\BIBentrySTDinterwordspacing}{\spaceskip=0pt\relax}
\providecommand{\BIBentryALTinterwordstretchfactor}{4}
\providecommand{\BIBentryALTinterwordspacing}{\spaceskip=\fontdimen2\font plus
\BIBentryALTinterwordstretchfactor\fontdimen3\font minus
  \fontdimen4\font\relax}
\providecommand{\BIBforeignlanguage}[2]{{%
\expandafter\ifx\csname l@#1\endcsname\relax
\typeout{** WARNING: IEEEtran.bst: No hyphenation pattern has been}%
\typeout{** loaded for the language `#1'. Using the pattern for}%
\typeout{** the default language instead.}%
\else
\language=\csname l@#1\endcsname
\fi
#2}}
\providecommand{\BIBdecl}{\relax}
\BIBdecl

\bibitem{qiu2014quantum}
J.~Qiu, ``Quantum communications leap out of the lab,'' \emph{Nature News},
  vol. 508, no. 7497, p. 441, 2014.

\bibitem{PhysRevA.98.030302}
I.~Cohen and K.~M\o{}lmer, ``Deterministic quantum network for distributed
  entanglement and quantum computation,'' \emph{Phys. Rev. A}, vol.~98, p.
  030302, Sep 2018.

\bibitem{chiribella2012optimal}
G.~Chiribella, ``Optimal networks for quantum metrology: semidefinite programs
  and product rules,'' \emph{New Journal of Physics}, vol.~14, no.~12, p.
  125008, 2012.

\bibitem{mccutcheon2016experimental}
W.~McCutcheon, A.~Pappa, B.~Bell, A.~Mcmillan, A.~Chailloux, T.~Lawson,
  M.~Mafu, D.~Markham, E.~Diamanti, I.~Kerenidis \emph{et~al.}, ``Experimental
  verification of multipartite entanglement in quantum networks,'' \emph{Nature
  communications}, vol.~7, no.~1, pp. 1--8, 2016.

\bibitem{briegel1998quantum}
H.-J. Briegel, W.~D{\"u}r, J.~I. Cirac, and P.~Zoller, ``Quantum repeaters: the
  role of imperfect local operations in quantum communication,'' \emph{Physical
  Review Letters}, vol.~81, no.~26, p. 5932, 1998.

\bibitem{PhysRevLett.71.4287}
M.~\ifmmode~\dot{Z}\else \.{Z}\fi{}ukowski, A.~Zeilinger, M.~A. Horne, and
  A.~K. Ekert, ````event-ready-detectors'' bell experiment via entanglement
  swapping,'' \emph{Phys. Rev. Lett.}, vol.~71, pp. 4287--4290, Dec 1993.

\bibitem{PhysRevA.72.042310}
A.~Sen(De), U.~Sen, i.~c.~v. Brukner, V.~Bu\ifmmode~\check{z}\else
  \v{z}\fi{}ek, and M.~\ifmmode~\dot{Z}\else \.{Z}\fi{}ukowski, ``Entanglement
  swapping of noisy states: A kind of superadditivity in nonclassicality,''
  \emph{Phys. Rev. A}, vol.~72, p. 042310, Oct 2005.

\bibitem{vedral2006introduction}
V.~Vedral, \emph{Introduction to quantum information science}.\hskip 1em plus
  0.5em minus 0.4em\relax Oxford University Press on Demand, 2006.

\bibitem{ji2022entanglement}
Z.~Ji, P.~Fan, and H.~Zhang, ``Entanglement swapping for bell states and
  greenberger--horne--zeilinger states in qubit systems,'' \emph{Physica A:
  Statistical Mechanics and its Applications}, vol. 585, p. 126400, 2022.

\bibitem{PhysRevLett.70.1895}
C.~H. Bennett, G.~Brassard, C.~Cr\'epeau, R.~Jozsa, A.~Peres, and W.~K.
  Wootters, ``Teleporting an unknown quantum state via dual classical and
  einstein-podolsky-rosen channels,'' \emph{Phys. Rev. Lett.}, vol.~70, pp.
  1895--1899, Mar 1993.

\bibitem{galindo2002information}
A.~Galindo and M.~A. Martin-Delgado, ``Information and computation: Classical
  and quantum aspects,'' \emph{Reviews of Modern Physics}, vol.~74, no.~2, p.
  347, 2002.

\bibitem{ji2019quantum}
Z.~Ji, H.~Zhang, H.~Wang, F.~Wu, J.~Jia, and W.~Wu, ``Quantum protocols for
  secure multi-party summation,'' \emph{Quantum Information Processing},
  vol.~18, no.~6, pp. 1--19, 2019.

\bibitem{PhysRevA.57.822}
S.~Bose, V.~Vedral, and P.~L. Knight, ``Multiparticle generalization of
  entanglement swapping,'' \emph{Phys. Rev. A}, vol.~57, pp. 822--829, Feb
  1998.

\bibitem{PhysRevLett.93.260501}
G.~Gour and B.~C. Sanders, ``Remote preparation and distribution of bipartite
  entangled states,'' \emph{Phys. Rev. Lett.}, vol.~93, p. 260501, Dec 2004.

\bibitem{song2014purifying}
W.~Song, M.~Yang, and Z.-L. Cao, ``Purifying entanglement of noisy two-qubit
  states via entanglement swapping,'' \emph{Physical Review A}, vol.~89, no.~1,
  p. 014303, 2014.

\bibitem{roa2021matching}
L.~Roa, T.~L. Purz, A.~Mu{\~n}oz, S.~Castro, G.~Hidalgo, and D.~Montoya,
  ``Matching for probabilistic entanglement swapping,'' \emph{arXiv preprint
  arXiv:2107.07689}, 2021.

\bibitem{huang2022entanglement}
C.-X. Huang, X.-M. Hu, Y.~Guo, C.~Zhang, B.-H. Liu, Y.-F. Huang, C.-F. Li,
  G.-C. Guo, N.~Gisin, C.~Branciard \emph{et~al.}, ``Entanglement swapping and
  quantum correlations via elegant joint measurements,'' \emph{arXiv preprint
  arXiv:2203.16207}, 2022.

\bibitem{branciard2010characterizing}
C.~Branciard, N.~Gisin, and S.~Pironio, ``Characterizing the nonlocal
  correlations created via entanglement swapping,'' \emph{Physical review
  letters}, vol. 104, no.~17, p. 170401, 2010.

\bibitem{nielsen2010quantum}
M.~A. Nielsen and I.~L. Chuang, ``Quantum computation and quantum
  information,'' 2010.

\bibitem{zangi2021combo}
S.~M. Zangi, J.-S. Wu, and C.-F. Qiao, ``Combo separability criteria and lower
  bound on concurrence,'' \emph{Journal of Physics A: Mathematical and
  Theoretical}, vol.~55, no.~2, p. 025302, 2021.

\bibitem{bennett1996mixed}
C.~H. Bennett, D.~P. DiVincenzo, J.~A. Smolin, and W.~K. Wootters,
  ``Mixed-state entanglement and quantum error correction,'' \emph{Physical
  Review A}, vol.~54, no.~5, p. 3824, 1996.

\bibitem{PhysRevA.57.1619}
V.~Vedral and M.~B. Plenio, ``Entanglement measures and purification
  procedures,'' \emph{Phys. Rev. A}, vol.~57, pp. 1619--1633, Mar 1998.

\bibitem{PhysRevA.75.052320}
J.~I. de~Vicente, ``Lower bounds on concurrence and separability conditions,''
  \emph{Phys. Rev. A}, vol.~75, p. 052320, May 2007.

\bibitem{lee2003convex}
S.~Lee, D.~P. Chi, S.~D. Oh, and J.~Kim, ``Convex-roof extended negativity as
  an entanglement measure for bipartite quantum systems,'' \emph{Physical
  Review A}, vol.~68, no.~6, p. 062304, 2003.

\bibitem{karimi2019measurability}
N.~Karimi, A.~Heshmati, M.~Yahyavi, M.~Jafarizadeh, and A.~Mohammadzadeh,
  ``Measurability of d-concurrence,'' \emph{Scientific reports}, vol.~9, no.~1,
  pp. 1--6, 2019.

\bibitem{PhysRevA.65.032314}
G.~Vidal and R.~F. Werner, ``Computable measure of entanglement,'' \emph{Phys.
  Rev. A}, vol.~65, p. 032314, Feb 2002.

\bibitem{PhysRevA.91.032327}
C.~Eltschka, G.~T\'oth, and J.~Siewert, ``Partial transposition as a direct
  link between concurrence and negativity,'' \emph{Phys. Rev. A}, vol.~91, p.
  032327, Mar 2015.

\bibitem{miranowicz2004comparative}
A.~Miranowicz and A.~Grudka, ``A comparative study of relative entropy of
  entanglement, concurrence and negativity,'' \emph{Journal of Optics B:
  Quantum and Semiclassical Optics}, vol.~6, no.~12, p. 542, 2004.

\bibitem{sabin2008classification}
C.~Sab{\'\i}n and G.~Garc{\'\i}a-Alcaine, ``A classification of entanglement in
  three-qubit systems,'' \emph{The european physical journal D}, vol.~48,
  no.~3, pp. 435--442, 2008.

\bibitem{rai2005negativity}
S.~Rai and J.~R. Luthra, ``Negativity and concurrence for two qutrits,''
  \emph{arXiv preprint quant-ph/0507263}, 2005.

\bibitem{tsujimoto2018high}
Y.~Tsujimoto, M.~Tanaka, N.~Iwasaki, R.~Ikuta, S.~Miki, T.~Yamashita, H.~Terai,
  T.~Yamamoto, M.~Koashi, and N.~Imoto, ``High-fidelity entanglement swapping
  and generation of three-qubit ghz state using asynchronous telecom photon
  pair sources,'' \emph{Scientific reports}, vol.~8, no.~1, pp. 1--6, 2018.

\bibitem{yu2004free}
C.-s. Yu and H.-s. Song, ``Free entanglement measure of multiparticle quantum
  states,'' \emph{Physics Letters A}, vol. 330, no.~5, pp. 377--383, 2004.

\bibitem{xiong2016multiple}
P.-Y. Xiong, X.-T. Yu, H.-T. Zhan, and Z.-C. Zhang, ``Multiple teleportation
  via partially entangled ghz state,'' \emph{Frontiers of Physics}, vol.~11,
  no.~4, pp. 1--8, 2016.

\bibitem{yan2010probabilistic}
F.~Yan and T.~Yan, ``Probabilistic teleportation via a non-maximally entangled
  ghz state,'' \emph{Chinese Science Bulletin}, vol.~55, no.~10, pp. 902--906,
  2010.

\bibitem{horodecki1999reduction}
M.~Horodecki and P.~Horodecki, ``Reduction criterion of separability and limits
  for a class of distillation protocols,'' \emph{Physical Review A}, vol.~59,
  no.~6, p. 4206, 1999.

\bibitem{quesada2012quantum}
N.~Quesada, A.~Al-Qasimi, and D.~F. James, ``Quantum properties and dynamics of
  x states,'' \emph{Journal of Modern Optics}, vol.~59, no.~15, pp. 1322--1329,
  2012.

\bibitem{uhlmann1976transition}
A.~Uhlmann, ``The “transition probability” in the state space of
  a-algebra,'' \emph{Reports on Mathematical Physics}, vol.~9, no.~2, pp.
  273--279, 1976.

\end{thebibliography}
\end{document}